\documentclass[lettersize,journal]{IEEEtran}
\usepackage{amsmath,amsfonts}
\usepackage{algorithmic}
\usepackage{algorithm}
\usepackage{multirow}
\usepackage{booktabs} 
\usepackage{array}    
\usepackage[caption=false,font=normalsize,labelfont=sf,textfont=sf]{subfig}
\usepackage{textcomp}
\usepackage{stfloats}
\usepackage{hyperref}
\usepackage{verbatim}
\usepackage{graphicx}
\usepackage{cite}
\usepackage{makecell}
\usepackage{threeparttable}
\usepackage{xcolor}

\UseRawInputEncoding
\begin{document}
	
\title{MAD: Meta Adversarial Defense Benchmark}

\author{
	Xiaoxu Peng$^{1}$, Dong Zhou$^{1}$, Guanghui Sun$^{1}$,~\IEEEmembership{Senior Member, IEEE}, Jiaqi Shi$^{1}$
 and Ligang Wu$^{1}$,~\IEEEmembership{Fellow, IEEE}
	\thanks{$^{1}$ X. Peng, D. Zhou, G. Sun, J. Shi, and L. Wu are with the Department of Control Science and Engineering, Harbin Institute of Technology, Harbin, China.	
	{\tt\footnotesize 21b904053@stu.hit.edu.cn}, {\tt\footnotesize dongzhou@hit.edu.cn}, {\tt\footnotesize guanghuisun@hit.edu.cn}, {\tt\footnotesize 22s104188@stu.hit.edu.cn}, {\tt\footnotesize ligangwu@hit.edu.cn}}
\thanks{This study was kindly supported by the National Key R\&D Program of China through grant 2019YFB1312001 and National Natural Science Foundation of China through Grant 62173107.}
\thanks{Manuscript received ; revised }}

\markboth{Journal of \LaTeX\ Class Files,~Vol.~14, No.~8, August~2021}%
{Shell \MakeLowercase{\textit{et al.}}: A Sample Article Using IEEEtran.cls for IEEE Journals}

\IEEEpubid{0000--0000/00\$00.00~\copyright~2021 IEEE}

\maketitle

\begin{abstract}
Adversarial training (AT) is a prominent technique employed by deep learning models to defend against adversarial attacks, and to some extent, enhance model robustness. However, there are three main drawbacks of the existing AT-based defense methods: expensive computational cost, low generalization ability, and the dilemma between the original model and the defense model. To this end, we propose a novel benchmark called meta adversarial defense (MAD). The MAD benchmark consists of two MAD datasets, along with a MAD evaluation protocol. The two large-scale MAD datasets were generated through experiments using 30 kinds of attacks on MNIST and CIFAR-10 datasets. In addition, we introduce a meta-learning based adversarial training (Meta-AT) algorithm as the baseline, which features high robustness to unseen adversarial attacks through few-shot learning. Experimental results demonstrate the effectiveness of our Meta-AT algorithm compared to the state-of-the-art methods. Furthermore, the model after Meta-AT maintains a relatively high clean-samples classification accuracy (CCA). It is worth noting that Meta-AT addresses all three aforementioned limitations, leading to substantial improvements. This benchmark ultimately achieved breakthroughs in investigating the transferability of adversarial defense methods to new attacks and the ability to learn from a limited number of adversarial examples. Our codes and attacked datasets address will be available at \url{https://github.com/PXX1110/Meta_AT}. 
\end{abstract}

\begin{figure}
	\centering	
	\includegraphics[scale=0.4]{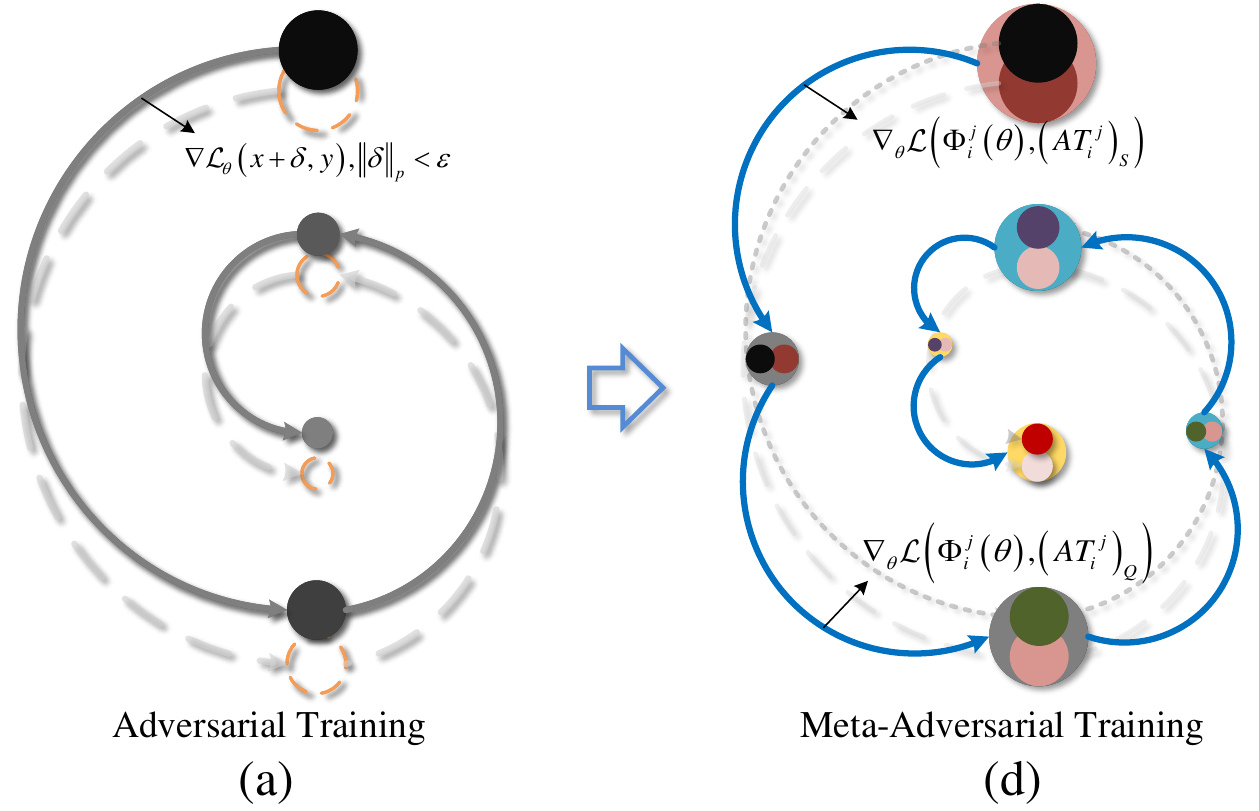}
	\caption{The diagram illustrates the updating of model parameters $\theta$ under different training methods. The directional arrows indicate the convergence direction of the parameters, while equations represent the associated gradients. In (a), the circles represent the current task parameters. In (b), the small circles with different colors within each task represent adversarial and clean examples, while the larger circles surrounding them represent the model parameters of the mini-AT. For comparison, the long dotted line represents conventional training, and the short dotted line represents AT.}
	\label{fig_1}
\end{figure}
\section{Introduction}
\IEEEPARstart{A}{dversarial} example is an important and interesting area of deep learning, generally referring to the fact that adding some small perturbations to a clean example can fool a well-trained model, and sometimes even affect the physical world. This poses a threat to cutting-edge applications such as autonomous driving\cite{wang2022review}, facial recognition\cite{bouchaffra2014nonlinear}, and intelligent healthcare\cite{zhou2020human}. As a result, researchers have focused on exploring different defense schemes to protect against such adversarial attacks, which have received significant attention. Existing adversarial defense methods can be broadly categorized into several types: 1) Model alteration\cite{akhtar2021advances}; 2) Detection as defence\cite{tao2018attacks, li2020connecting}; 3) Input transformation\cite{guo2017countering, raff2019barrage}; 4) Certified  defences\cite{zhai2020macer, jia2019certified}; 5) Other defences\cite{cemgil2019adversarially, he2020defending}. 

AT\cite{madry2017towards} is a widely used framework in Model alteration. It is acknowledged as one of the strongest principled defenses against adversarial attacks and has inspired the development of many other adversarial defense methods. 
\IEEEpubidadjcol
However, AT suffers from certain drawbacks, including high computational cost, low generalization performance against new attacks, and degradation of the classification performance of the original model.

To mitigate the computational burden associated with AT, Shafahi \textit{et al}.\cite{shafahi2019adversarial} addressed the issue by leveraging the gradient information computed during parameter updates, eliminating the need for costly generation of adversarial examples. Their proposed algorithm, termed ``Free" adversarial training, incurs minimal additional overhead compared to natural training and can achieve a speedup of 7 to 30 times compared to the original AT. To improve the generalization ability of AT, Maini \textit{et al}.\cite{maini2020adversarial} considered the worst-case scenario in the steepest direction of the entire model as an extension of standard project gradient descent (PGD), combining several perturbations into a single attack. The experimental result shows that the trained classifier can combat the bounded perturbations of the ``1", ``2" and ``$\infty$" norms. To preserve the original classification ability of AT, Farnia \textit{et al}.\cite{farnia2018generalizable} introduced an AT regularization technique utilizing spectral normalization. This method improves the classification performance of the model while maintaining its defensive capabilities. 

After continuous research and exploitation, it has been found that the defense effectiveness of AT is highly coupled with the adversarial attacks involved in the training samples. That is, the trained model only has strong defense capabilities against the attacks used in the training process. It remains an ongoing challenge to develop adversarial defense methods that can withstand new attack techniques.\par
\begin{figure*}
	\centering
	\includegraphics[scale=1.2]{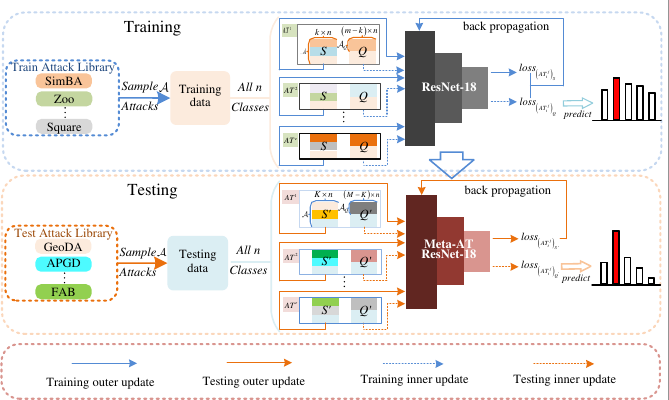}
	\caption{The details of \textit{``$\mathcal{A}$-way, K-shot"} Meta-AT. Different attacks in the attack library are represented by different colors, and the adversarial examples in \textit{S}, \textit{Q}, \textit{S'} and \textit{Q'} have the same color as their corresponding attacks. ResNet-18 is taken as an example for the target network.}
	\label{fig_2}
\end{figure*}

 
To this end, we propose a MAD benchmark. This benchmark includes two large-scale MAD datasets: MAD-M and MAD-C, along with a MAD evaluation protocol. The MAD datasets are constructed by 30 mainstream adversarial attacks on the MNIST\cite{lecun1998gradient} and CIFAR-10\cite{krizhevsky2009learning}. In MAD protocol, we explain the usage of MAD datasets, the training pattern of the backbone network, and introduce a novel metric called equilibrium defense success rate (EDSR), which provides a comprehensive and unbiased evaluation of defense method effectiveness. Meta-AT is the baseline algorithm for MAD that features high robustness to unseen adversarial attacks after few-shot learning. As depicted in Fig.~\ref{fig_1}, we illustrate the basic principles of model parameter update methods used in AT and Meta-AT. It can be seen that the parameter update approach for AT is similar to conventional training. However, in Meta-AT, we transform the one-step update into internal and external updates. The AT is treated into multiple mini-ATs as ``base" tasks, enabling the model to learn from diverse adversarial examples of fixed attack types in each epoch. Furthermore, the detailed procedure for Meta-AT is depicted in Fig.~\ref{fig_2}.\par 
Experimental results demonstrate that Meta-AT achieves a maximum EDSR of 99.77\% for MAD-M and 92.98\% for MAD-C. When evaluating the performance of the defended models on the original clean examples, only a slight decrease is observed, with a maximum decrease of 0.06\% for MNIST and 6.24\% for CIFAR-10. Therefore, in comparison to other state-of-the-art (SOTA) adversarial defense methods such as AT, Adversarial Training with Transferable Adversarial Examples (ATTA)\cite{zheng2020efficient}, and You Only Propagate Once (YOPO)\cite{zhang2019you}, Meta-AT achieves a remarkably high meta-defense capability based on the evaluation criteria of the MAD benchmark. It is worth noting that the MAD benchmark effectively addresses these three aforementioned shortcomings of AT, learns from examples across different domains, and mitigates most of the potential ``blind spots" in the model. In summary, our specific contributions can be summarized as follows:
\begin{itemize}
	\item 
	We present the first comprehensive MAD benchmark, which includes two extensive MAD datasets, a MAD evaluation protocol, and a Meta-AT baseline algorithm.
	\item
	The two MAD datasets, MAD-M and MAD-C, are constructed by 30 mainstream adversarial attack methods based on the MNIST and CIFAR-10. Additionally, we propose a versatile evaluation metric called EDSR in the MAD evaluation protocol, providing a comprehensive and unbiased assessment of defense method effectiveness.
	\item
	We proposed a Meta-AT features high robustness to unseen adversarial attacks through few-shot learning. It achieves the highest EDSR score compared to the SOTA adversarial defense methods while maintaining excellent CCA. 
\end{itemize}

The remainder organizations of this article are: In Section 2, related works are introduced. In Section 3, we present the basic principles of our proposed MAD benchmark. Experiments and results are given in Section 4. The conclusion is provided in Section 5.

\section{Related Works}
In this section, we will provide an overview of the fundamental concepts involved in this paper and existing adversarial defense methods related to meta-learning.
\subsection{Adversarial Training}
The adversarial defense method based on AT is often regarded as a simple and rudimentary approach for countering adversarial attacks and once became the occupation standard method. The basic operation is to train clean examples and adversarial examples generated by certain (or several) attacks together. In this way, the model can learn to be robust to adversarial input. Madry\cite{madry2017towards} transforms AT into a non-convex and non-concave saddle-point problem, fitting it to the min-max problem as shown in Eq.~\ref{deqn_ex1}. Among Eq.~\ref{deqn_ex1}, inner maximization corresponds to an arbitrary adversarial attack. They argue that the greater the intensity of the adversarial attack, the stronger the defense capability of the model trained with it.
\begin{equation}
	\label{deqn_ex1}
	\min _\theta \mathbb{E}_{(\mathbf{x}, y) \sim \mathcal{D}}\left[\max _{\|\delta\|_p<\epsilon} \mathcal{L}_\theta(\mathbf{x}+\delta, y)\right]
\end{equation}

where $\mathcal{L}_\theta(\mathbf{x}+\delta, y)$ is the loss function of the network with parameter $\theta$. (x, y) is the training pair sampled from the training set $\mathcal{D}$. $\delta$ is an adversarial perturbation, whose $p$ norm is smaller than the parameter $\epsilon$. According to Eq.~\ref{deqn_ex1}, the mean idea of AT is to expect to minimize the network parameters with the lowest loss under the class label while maximizing the adversarial perturbation. In addition to the adversarial defense methods mentioned earlier, there are other improved approaches for AT such as virtual AT\cite{miyato2018virtual}, integrated AT\cite{meng2023integrating}, and stability AT\cite{ke2019araml}.

\subsection{Meta-Learning Related Adversarial Defense Methods}
\begin{figure}[t]
	\centering	
	\includegraphics[scale=1.2]{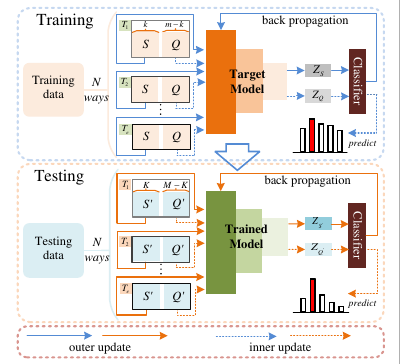}
	\caption{The meta-learning procedure for the \textit{``N-way, K-shot"} classification task within one epoch is illustrated. The testing sets are denoted as $S'$ and $Q'$. Here, \textit{k}, \textit{m-k}, \textit{K}, and \textit{M-K} represent the number of sampled examples for each respective set. The variable \textit{Z} represents the classification feature.}
	\label{fig_3}
\end{figure}

Meta-learning has become a popular and effective approach for learning tasks with limited data. The primary objective of meta-learning is to train a model that can rapidly adapt to new tasks using only a small amount of available data. In few-shot classification tasks, the most common scenario is to address the \textit{``N-way, K-shot"} problem. The detailed procedure for it is shown in Fig.~\ref{fig_3}. One prominent method in this field is model-agnostic meta-learning (MAML)\cite{wang2021visual}. Yu \textit{et al.}\cite{yu2023pid} proposed a FracMAML, which utilized fractional-order optimization to improve the MAML algorithm. It aims to address the problem of oscillation and difficult convergence that often occurs during the later stages of training.

The integration of meta-learning with adversarial examples is currently a vibrant and active area of research. Goldblum \textit{et al.}\cite{goldblum2020adversarially} developed the Adversarial Query (AQ) algorithm by substituting the query set in few-shot classification tasks with adversarial examples. AQ is a well-known approach that enhances the robustness of few-shot learning against adversarial examples through AT. Liu \textit{et al.}\cite{liu2021long} proposed a Long-Term Cross-Adversarial Training (LCAT) method, which achieves similar effects to AQ but requires only half the AT time. The main contribution of LCAT is the cross-utilization of AT to reduce the computational cost in the AQ algorithm. Qi \textit{et al.}\cite{qi2022cross} introduced a meta-learning-based AT framework (MBATF), which includes attention mechanisms, encoders, and decoders. By incorporating adversarial examples into the few-shot classification task, MBATF shows promising results in mitigating the cross-domain problem in few-shot learning. The core of these similar related works are to utilize adversarial defense methods to enhance the ability of few-shot learning models to deal with adversarial examples, or thereby improving model robustness.

Several researchers have proposed innovative frameworks that leverage meta-learning for improved defense against adversarial attacks. Ma \textit{et al.}\cite{ma2019metaadvdet} introduced a dual-network framework based on meta-learning. Their approach involves two networks that iteratively learn a robust attack detection model and utilize limited examples to effectively detect new adversarial attacks. This work contributes significantly to the field of Detection as defense. Metzen\textit{et al.}\cite{metzen2021meta} utilizes meta-learning to conduct adversarial training on various meta-patches generated based on IFGSM, thus generating models that can defend against universal patch attacks. This approach has made contributions towards defending against universal patch attacks in a targeted manner. It is important to emphasize that our focus is on AT within classification tasks, rather than few-shot classification tasks or enhancing robustness for few-shot classification tasks. The primary objective is to train a classification model with high accuracy, capable of defending against known attacks, and quickly adapting to defend against unknown attacks. These approaches contribute to the growing body of knowledge in the field of meta-learning techniques to improve adversarial defense mechanisms and offer valuable insights for our advancements in adversarial defense strategies. 
\begin{table*}[h]
	\caption{Adversarial Algorithm.\label{table_1}}	
	\centering
	\begin{threeparttable}
		\begin{tabular*}{1.0\linewidth}{@{}lllll@{}}	
			\toprule[1pt]
			ID & Name & Measurement & Knowledge & Introduction\\
			\midrule[0.5pt]	
			0&\makecell[l]{JSMA\cite{papernot2016limitations}}  & $L_2$ & White-box & \makecell[l]{Constructing an adversarial saliency map involves generating input features that have the most\\significant influence on the output.}\\	
			
			1&\makecell[l]{Deep-Fool\cite{moosavi2016deepfool}} & $L_2$ & White-box & \makecell[l]{Calculating the minimum distance between normal samples and the decision boundary of the\\ model in order to generate perturbations.}\\
			
			2&\makecell[l]{Universal-\\Perturbation\cite{moosavi2017universal} } & $L_\infty$& White-box & \makecell[l]{Searching for general perturbations at training points, aggregate perturbation vectors that send\\ successful data points to decision boundaries.}\\
			
			3&\makecell[l]{Newton-Fool\cite{jang2017objective} }& $L_2$/$L_0$& White-box & \makecell[l]{Generating adversarial samples based on model gradient.}\\
			
			4&\makecell[l]{Boundary-\\Attack\cite{brendel2017decision} }& $L_2$& Black-box & \makecell[l]{Reducing the size of disturbances gradually starting fromlarge disturbances while maintaining\\aggressiveness based on decision boundary.}\\
			
			5&\makecell[l]{Elastic Net\cite{chen2018ead}} &  $L_1$ & White-box & \makecell[l]{Expressing the adversarial attack problem as the elastic network regularization optimization\\problem.}\\
			
			6&\makecell[l]{Zoo-Attack\cite{chen2017zoo}}  & $L_0$ & Black-box & \makecell[l]{Generating adversarial samples based on the gradient of the zero-order optimization estimation\\target model.}\\
			
			7&\makecell[l]{Spatial-Trans-\\formation\cite{engstrom2019exploring}}& $L_N$ & Black-box & \makecell[l]{Generating perturbation based on natural perturbation categories such as translation.}\\
			
			8&\makecell[l]{Hop-Skip-\\Jump\cite{chen2020hopskipjumpattack}}&$L_\infty$/$L_2$ & Black-box & \makecell[l]{Generating perturbation gradient direction estimation at decision boundaries based on binary\\information.}\\
			
			9&\makecell[l]{Sim-BA\cite{guo2019simple}}&$L_2$ & Black-box & \makecell[l]{Adding sample vector from a predefined orthogonal offset on the original image.}\\
			
			10&\makecell[l]{Shadow-\\Attack\cite{ghiasi2020breaking}}&$L_N$ & White-box & \makecell[l]{Keeping the adversarial sample far away from the decision-making boundary under the premise\\that itis not perceivable.}\\
			
			11&\makecell[l]{GeoDA\cite{rahmati2020geoda}}& $L_\infty$ & Black-box & \makecell[l]{Black-box iterative attack algorithm based on query.}\\
			
			12&\makecell[l]{Wasserstein\cite{wong2019wasserstein}}&$L_N$ & White-box & \makecell[l]{Searching for adversarial perturbations of Wasserstein distance based on Sinkhorn iteratively.}\\
			
			13&\makecell[l]{FGSM\cite{goodfellow2014explaining}}& $L_\infty$ & White-box & \makecell[l]{Adding perturbations in reverse based on the gradient direction of normal samples.}\\
			
			14&\makecell[l]{BIM\cite{kurakin2018adversarial}}& $L_\infty$ & White-box & \makecell[l]{Adversarial sample attacks against the physical world.}\\
			
			15&\makecell[l]{CW\cite{carlini2017towards}}&$L_\infty$ & White-box & \makecell[l]{Finding the least loss function additive perturbation term that causes the neural network\\to misclassify, and transform the problem into a convex optimization problem.}\\
			
			16&\makecell[l]{MIFGSM\cite{dong2018boosting}} & $L_\infty$ & White-box & \makecell[l]{Generating perturbations by an iterative approach based on momentum to find\\counter perturbations.}\\
			
			17&\makecell[l]{TIFGSM\cite{dong2019evading}}& $L_\infty$ & White-box & \makecell[l]{Generating a more transferable perturbations byoptimizing the perturbation of the image\\transformation set.}\\
			
			18&\makecell[l]{PGD\cite{madry2017towards}} & $L_\infty$ & White-box & \makecell[l]{Generating adversarial perturbation based on gradient projection direction iterative algorithm.}\\
			
			19&\makecell[l]{PGD-L2\cite{madry2017towards}}& $L_2$ & White-box & \makecell[l]{Generating L2 perturbation based on gradient projection direction iterative algorithm.}\\
			
			20&\makecell[l]{TPGD\cite{zhang2019theoretically}}& $L_\infty$ & White-box & \makecell[l]{Generating adversarial perturbations by an iterative algorithm of gradient projection\\directionbased on KL-Divergence loss.}\\
			
			21&\makecell[l]{RFGSM\cite{tramer2017ensemble}}& $L_\infty$ & White-box & \makecell[l]{A single-step gradient attack method based on small random step size.}\\
			
			22&\makecell[l]{APGD\cite{croce2020reliable}}&$L_\infty$/$L_2$ & White-box & \makecell[l]{Generating adversarial perturbation an iterative algorithm of gradient projection direction based\\on variable step size with loss ``ce".}\\
			
			23&\makecell[l]{APGD2\cite{croce2020reliable}}&$L_\infty$/$L_2$ & White-box & \makecell[l]{Generating adversarial perturbation an iterative algorithm of gradient projection direction based\\ on variable step size with loss ``dlr".}\\
			
			24&\makecell[l]{FFGSM\cite{wong2020fast}}&$L_\infty$ & White-box & \makecell[l]{Gradient single-step attack method based on random initialization.}\\
			
			25&\makecell[l]{Square\cite{andriushchenko2020square}}&$L_\infty$/$L_2$ & Black-box & \makecell[l]{An iterative algorithm to find feasible set boundaries against perturbations by updating local\\squares at random locations.}\\
			
			26&\makecell[l]{TIFGSM2\cite{dong2019evading}}& $L_\infty$ & White-box & \makecell[l]{Generating a more transferable perturbations byoptimizing the perturbation of the image\\transformation set with different resize rate.}\\
			
			27&\makecell[l]{EOTPGD\cite{liu2018adv}}& $L_\infty$ & White-box & \makecell[l]{An iterative method for generating adversarial samples by estimating the direction of gradient\\projection by multiple random vectors.}\\
			
			28&\makecell[l]{One-Pixel\cite{su2019one}}& $L_0$ & Black-box & \makecell[l]{Generating single-pixel perturbations based on differential evolution.}\\
			
			29&\makecell[l]{FAB\cite{croce2020minimally}}& $L_\infty$/$L_2$/$L_1$ & White-box & \makecell[l]{Generating minimal perturbation based on fast adaptive boundaries.}\\
			\bottomrule[1pt]
		\end{tabular*}	
		\begin{tablenotes}
			\footnotesize
			\item $L_N$ indicates the unconventional perturbation measurement
		\end{tablenotes}
	\end{threeparttable}
\end{table*}
\section{MAD Benchmark}
We will provide a detailed introduction to the MAD benchmark, covering three main aspects: MAD datasets, MAD evaluation protocol, and the Meta-AT baseline algorithm.
\subsection{MAD Datasets}
To facilitate experimental analysis and subsequent innovation exploitation, we conducted 30 attacks on the MNIST and CIFAR-10 datasets using a ResNet-18 model trained on clean examples. The details of these attacks, along with their corresponding IDs, can be found in table~\ref{table_1}. Attacking both the training and test examples in traditional datasets is time-consuming and not significant for the few-shot learning pattern. Therefore, we select the test data from MNIST and CIFAR-10 for the attacks and obtain a large set of attacked datasets. The obtained attacked datasets MAD-M and MAD-C are composed of different seed attacked datasets. 

\begin{figure}
	\centering	
	\subfloat[]{\includegraphics[scale=0.267]{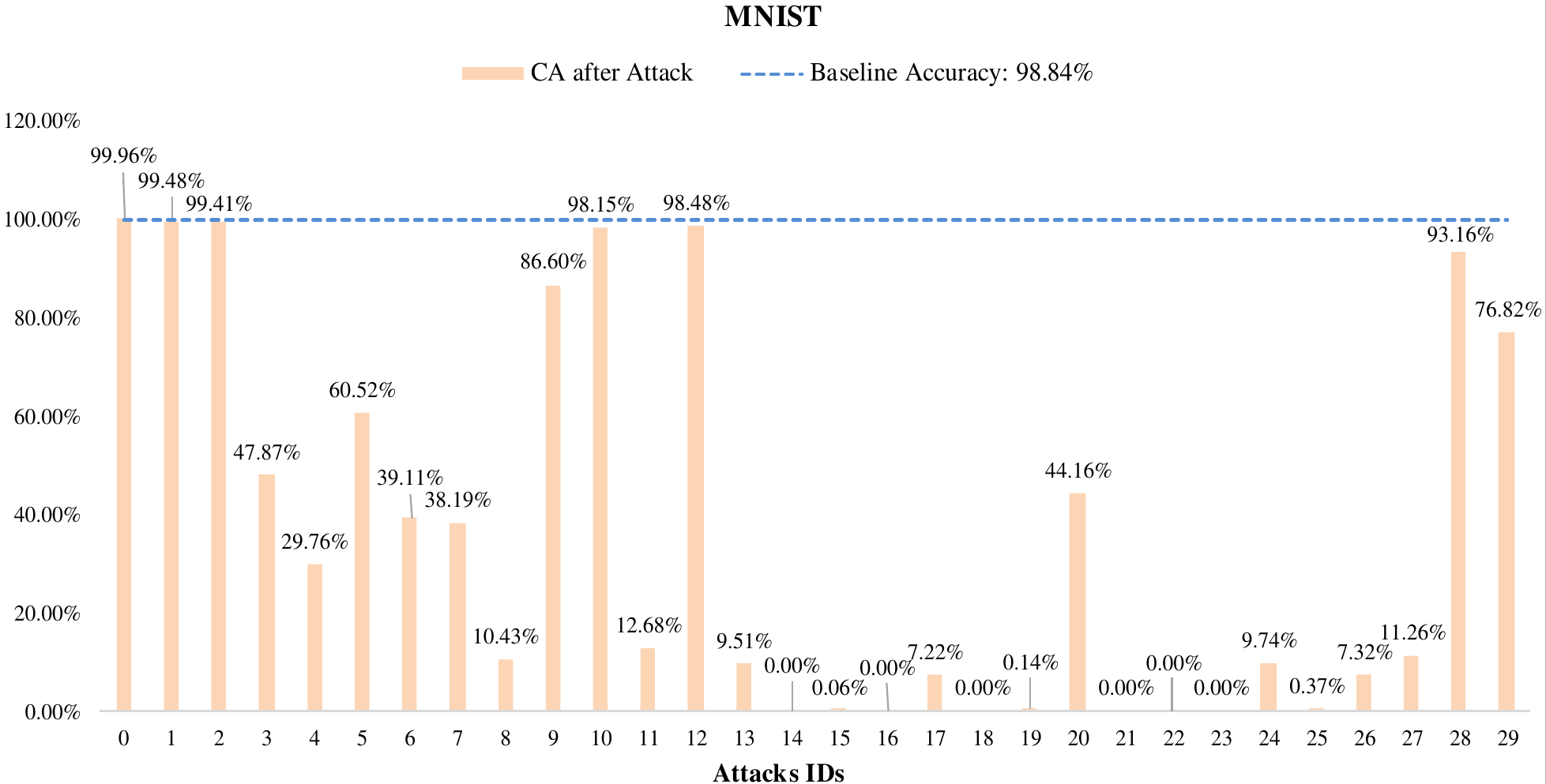}
		\label{fig_first_case}}
	\hfil
	\subfloat[]{\includegraphics[scale=0.276]{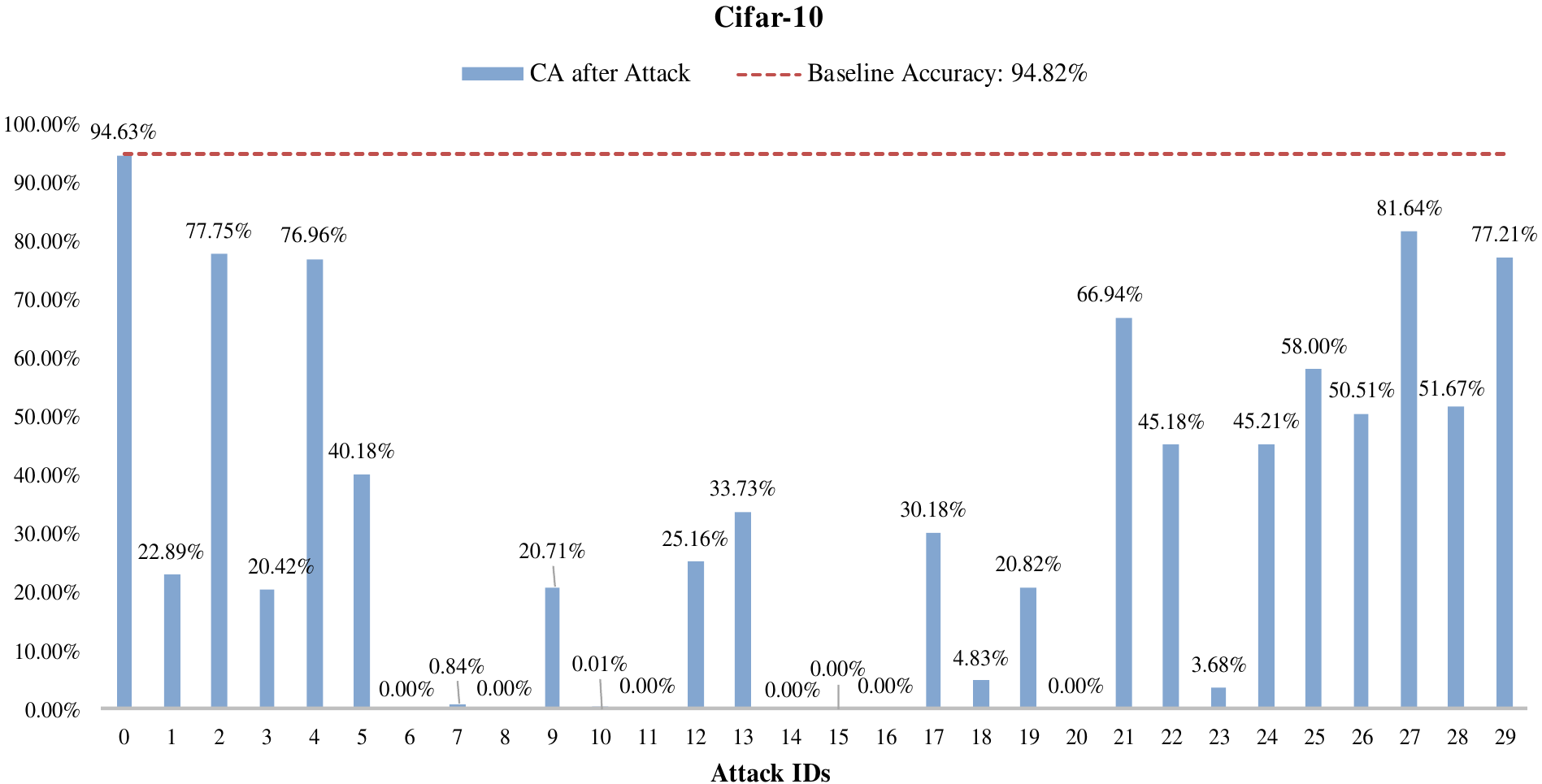}
		\label{fig_second_case}}
	\caption{CAs of the two initial test dataset after various attacks. The CCAs indicated by the dashed lines, are taken as the baselines.}
	\label{fig_4}.
	
\end{figure}

At the beginning of the dataset creation, we selected attacks from two well-known and simple attack libraries, Adversarial Robustness Toolbox (ART) and torchattack attack libraries. The CCAs of the model in the original test dataset are MNIST: 98.84\% and CIFAR-10: 94.82\%. To maintain the model classification balance, we included all classes in datasets for training and testing. No additional reinforcement operations were performed on the clean model, and the simplest training method was used. The CAs of the original test dataset after various attacks are given by the histogram in Fig.~\ref{fig_4}. As can be seen from the table, not all attacks were effective, and their effectiveness varied across datasets. Generally, larger image sizes in the dataset resulted in more pronounced attack effects. Some attacks did not have sufficient examples, so specific attacks were removed from the experiments. Therefore, the 0, 1, 2, 9, 10, 12, 24, and 28 attacks in MAD-M and the attack 0 in MAD-C were removed during the experiments. Among the attacks, the 24th attack in the eighth class stands out as it has a very low number of adversarial examples and resulted in an uneven distribution of examples among the categories. Due to this imbalance, we had to abandon this attack.

\begin{figure}[t]
	\centering
	\includegraphics[scale=0.7]{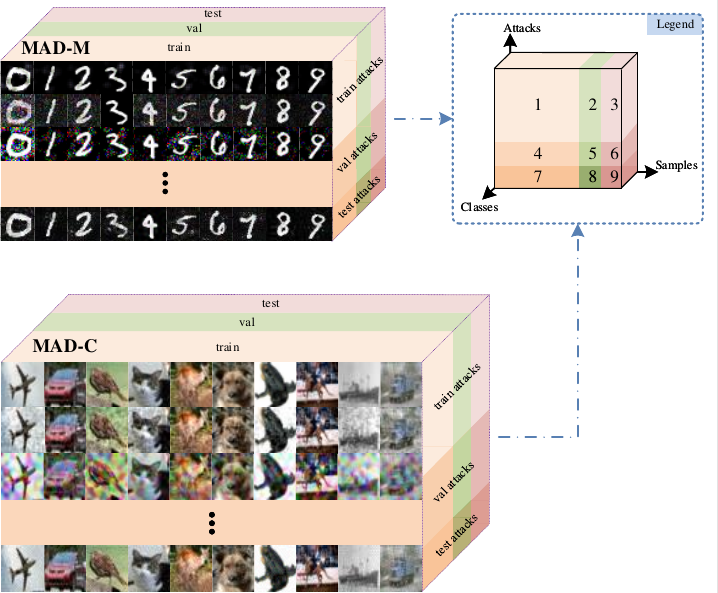}
	\caption{The details and simple examples of MAD-M and MAD-C.}
	\label{fig_5}
\end{figure}

Careful screening was applied to retain only successfully attacked examples under each attacked dataset. This resulted in each attacked dataset having a CA of 0\% on the initially trained clean model. MAD-M consists of 10 classes with a single image size of $28\times28$ under 22 attacks, while MAD-C consists of 10 classes with a single image size of $32\times32$ under 29 attacks. It can be seen that the datasets we used are more comprehensive. The training, validation, and test sets were divided based on the partitioning pattern of few-shot learning, taking into account the year and perturbation measurements proposed by different attacks. For better evaluation, examples from the original test dataset and the MAD datasets were split in a $3:1:1$ ratio for each class. Fig.~\ref{fig_5} shows the two types of MAD datasets obtained, with legends added to improve readability and usability. 

\begin{figure}[t]
	\centering
	\includegraphics[scale=0.5]{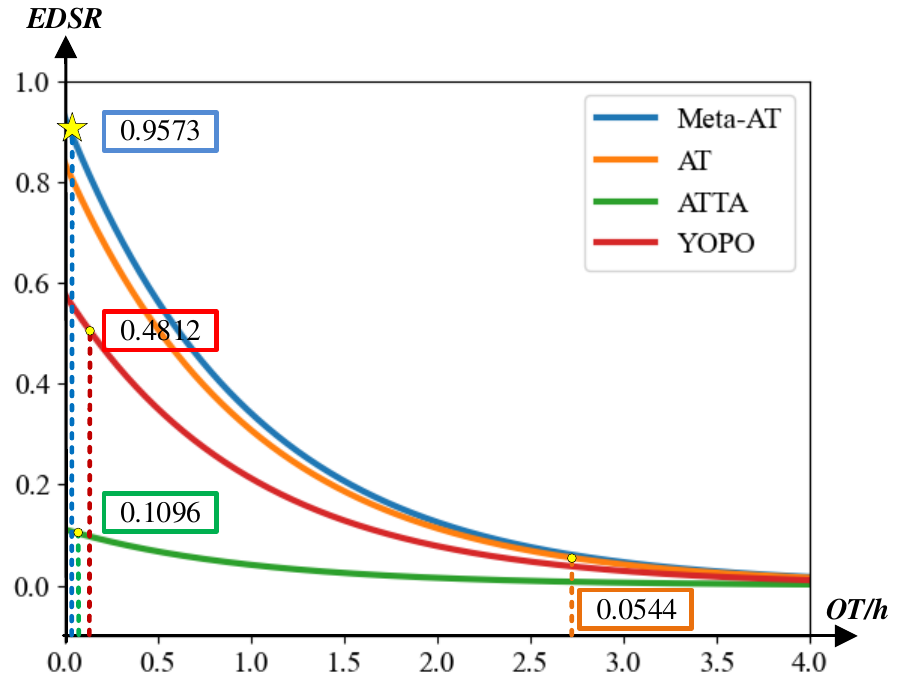}
	\caption{The EDSRs of different defense methods under the PGD attack.}
	\label{fig_6}
\end{figure}
\subsection{MAD Evaluation Protocol}
\subsubsection*{\bf Dataset setting}
There are different ways of using the MAD datasets. More specifically, ``1+2" is used for Meta-AT training, and ``4+5" is used for Meta-AT validation. ``7+8" is used to fine-tune the model previously trained by Meta-AT to defend against unknown attacks. ``3, 6" is used to test Meta-AT trained model defense against learned attacks, and ``9" is used to test Meta-AT trained model defense against unknown attacks. To ensure the fairness of the experiment, the AT-based algorithms used for comparison use the same datasets. More specifically, ``1+4" is used for AT training, and ``2+5" is used for AT validation. ``9" is used to test the AT defense against unknown attacks. ``3, 6" is used to test the AT defense against learned attacks. 

\subsubsection*{\bf Evaluation metrics}
To showcase the effectiveness of the proposed Meta-AT approach, the performance can be assessed and compared using the following four metrics: CA, Operating Time (OT), Defense Success Rate (DSR), and EDSR. The CA, OT, and DSR are conventional metrics, whereas EDSR is a novel and comprehensive metric specifically introduced for Meta-AT. CA measures the accuracy of the models in correctly classifying both clean samples and adversarial examples under different scenarios. OT quantifies the time required for the model to reach its highest defense rate, indicating the efficiency and speed of the Meta-AT approach compared to other AT-based techniques. DSR can be obtained by observing the amplitude of CA increase before and after defense. The basic formula for DSR is shown in Eq.~\ref{deqn_ex2}.
\begin{equation}
	\label{deqn_ex2}
	DSR=\frac{{CA}_{\mathcal{D}}-{CA}_{\mathcal{A}}}{{CCA}-{CA}_{\mathcal{A}}}
\end{equation}

where ${CA}_{\mathcal{D}}$ and ${CA}_{\mathcal{A}}$ represent the CA of the model after defense and after attack, respectively.

Fig.\ref{fig_6} shows an example of EDSRs for different adversarial defense methods under the PGD attack. When the x-axis value of OT is set to 0, the corresponding y-axis values indeed represent the DSRs of different defense algorithms. It can be observed that the AT outperforms ATTA and YOPO, but when efficiency is also considered, the AT algorithm performs the worst. Therefore, to comprehensively and fairly evaluate the effectiveness of AT-based methods, we introduced the innovative EDSR metric, which combines accuracy and efficiency. A high EDSR indicates that the model can achieve significant defense capability with minimal training effort and in a short period of time. The specific formula of EDSR is shown in Eq.~\ref{deqn_ex3}. 
\begin{equation}
	\label{deqn_ex3}
	EDSR= DSR \times e^{-{OT}}
\end{equation}
\subsection{Meta-AT baseline algorithm}
The basic configuration of Meta-AT typically involves the following elements:
\begin{itemize}
	\item 
	\textbf{Task \textit{AT}:} The dataset is partitioned into multiple meta-AT-tasks \textit{$\{{AT}^j\}$}, which follow a certain data distribution $\mathcal{P}$.
	\item
	\textbf{Support set (\textit{S}) and Query set (\textit{Q}):} Each meta-AT-task consists of an attacked \textit{S} and an attacked \textit{Q}. The attacked \textit{S} is used to fine-tune the parameters of the model in the inner loop, while the attacked \textit{Q} is utilized to assess the performance of the model in the outer loop.
	\item
	\textbf{Way:} The term ``way" in Meta-AT refers to the number of fixed attack categories, denoted as $\mathcal{A}$, that each meta-task learns. This notion is analogous to the concept of ``way" in traditional classification tasks, where it represents the number of fixed classes \textit{N} that a model learns to classify.
	\item
	\textbf{Shot:} It denotes the number of examples sampled for each ``way" in the \textit{Q}.
	\item
	\textbf{Episode (\textbf{e}):} It refers to the number of meta-AT-tasks that the entire task is divided into at each epoch.
\end{itemize}

The overall details of Meta-AT are given in Fig.~\ref{fig_2}. To enhance readability, we provide a detailed explanation of it. Specifically, take the training progress as an example: The meta-task of each batch is a mini-AT. Where \textit{S} is composed of adversarial examples generated using a set of $\mathcal{A}$ attacks, combined with clean examples. The \textit{Q} is composed of adversarial examples generated by $\mathcal{A}_Q$ attacks which are selected from $\mathcal{A}$ attacks used in \textit{S}, along with clean examples. That is, the types of attacks in \textit{S} and \textit{Q} cross during training. Data balancing is performed by selecting a specific number of adversarial examples under all \textit{n} classes per attack ($k \times n$) for \textit{S} and a specific number of adversarial examples per attack for \textit{Q} ($(m-k) \times n$), ensuring that the adversarial examples of \textit{S} and \textit{Q} do not intersect. The objective is to minimize the classification loss on both \textit{S} and \textit{Q}. The overall loss used for backpropagation includes the classification loss for both sets. The CA on \textit{Q} serves as a measure of the overall model's CA and defense ability after learning through several mini-ATs.

During the testing or validation phase: The test sets are represented as \textit{S'} and \textit{Q'}. These sets do not overlap and are used to learn the model's ability to defend against unknown attacks. The adversarial and clean examples in \textit{S'} and \textit{Q'} are selected from $(K \times n \times \mathcal{A})$ examples and$((M-K) \times n \times \mathcal{A}_Q)$ examples under all \textit{n} classes, respectively. Such a Meta-AT task can be analogous to the few-shot learning classification task called \textit{``$\mathcal{A}$-way, K-shot"}. Testing progress focuses on observing how well the model adapts quickly to new attacks. Therefore, only the data of \textit{S'} is used to fine-tune the network, and verified with \textit{Q'}. The CA on \textit{Q'} is selected to measure the EDSR. 

More concretely, the optimized objective of Meta-AT is shown in Eq.~\ref{deqn_ex4}. 
\begin{equation}
	\label{deqn_ex4}
	\min _\theta \underset{{AT}_a \sim P(AT)}{\mathrm{E}}\left[\max _{\|\delta_b\|_p<\epsilon} \mathcal{L}\left(\Phi_{\phi\left(\theta, {AT}_a\right)}(x+\delta_b), y\right)\right]
\end{equation}

Note that $\Phi(\theta)$ is a model represented by a parameterized function. We expect to obtain the model parameter $\theta$ at the time of minimal loss as the fine-tuning model $\phi(\theta, {AT}_a)$ adapts to the new AT task under the $b_{th}$ attack. The steps of the Meta-AT training method are shown in Algorithm~\ref{alg:alg1}.

\begin{algorithm}[t]
	\caption{Meta-Adversarial Training}\label{alg:alg1}
	\begin{algorithmic}
		\STATE 
		\STATE \textbf{Require:} Base model $\Phi(\theta)$, fine-tuning algorithm $\phi(\theta)$, learning rate $\lambda$, $\beta$, the number of adversarial attacks $\mathcal{A}$, the number of episodes $e$, and distribution over AT tasks $P(AT)$;
		\STATE Initialize all parameters;
		\STATE \textbf{while} not done \textbf{do}
		\STATE \hspace{0.5cm}Sample batch of tasks, $\left\{{AT}_i\right\}_{i=1}^\mathcal{A}$, where $A T_i=\left\{\left({AT}_i^j\right)_S,\left({AT}_i^j\right)_Q \mid j \in[1, e]\right\}$
		\STATE \hspace{0.5cm}\textbf{for} $i = 1,...,\mathcal{A}$ \textbf{do}
		\STATE \hspace{1cm}\textbf{for} $j = 1,..., e$ \textbf{do}
		\STATE \hspace{1cm}Fine-tune model $\phi$ on task $({AT}_i^j)_S$. 
		\STATE \hspace{1cm}$\theta_i^{j^{\prime}}=\theta-\beta \nabla_\theta \mathcal{L}\left(\phi(\theta),\left({AT}_i^j\right)_S\right)$.
		\STATE \hspace{1cm}Compute gradient $g_i^j=\nabla_\theta \mathcal{L}\left(\Phi_i^j(\theta_i^{j^{\prime}}),\left({AT}_i^j\right)_Q\right)$.
		\STATE \hspace{1cm}\textbf{end for}
		\STATE \hspace{0.5cm}\textbf{end for}
		\STATE \hspace{0.5cm}Update base model parameters: 
		\STATE \hspace{0.5cm}$\theta \leftarrow \theta-\frac{\lambda}{e} \sum g_i^j$
		\STATE \textbf{end while}
	\end{algorithmic}
	\label{alg1}
\end{algorithm}

\subsubsection*{\bf Why Meta-AT works?}
Splitting the optimization problem into multi-step learning can help prevent overfitting and promote broader learning. This approach has been successfully demonstrated in various examples. Where the progressive networks\cite{rusu2016progressive} effectively simplify subtasks before tackling more complex subtasks. By sequentially learning subtasks and building upon previous knowledge, models can avoid catastrophic forgetting and improve efficiency and effectiveness in learning. The PGD attack, originating from the Fast Gradient Sign Method (FGSM) attack, is an adversarial attack technique that performs multiple iterations within the model. By iteratively updating the input to maximize the loss or minimize the accuracy of the model, the PGD attack aims to achieve a stronger attack effect compared to the FGSM attack. This iterative approach can help to find more effective adversarial examples. Meta-learning involves decomposing a complex task into multiple smaller tasks, allowing the model to fine-tune itself on these smaller tasks. In this way, meta-learning enables the learning of model parameters that are more adaptable to changes in tasks, thus improving transferability across different domains. This strategy facilitates better generalization and can lead to improved performance in different domains. Similarly, our Meta-AT approach transforms a single AT task into multiple smaller AT tasks. By training the model on these numerous smaller tasks, the model can comprehensively learn various vulnerabilities or ``blind spots" and bridge the gaps between different adversarial domains. As a result, Meta-AT not only improves the model's robustness against different types of attacks but also enhances the model's accuracy on its primary tasks.

\begin{table}[t]
	\caption{Parameters of Meta-AT.\label{table_2}}
	\centering
	\begin{threeparttable}
		\begin{tabular}{lll}
			\toprule[1pt] 
			\textbf{Parameters}  &   \textbf{Default} &     \textbf{Description}                 \\
			\midrule[0.5pt]	
			$\beta$         &            0.01          &  Inner update learning rate.     \\
			$\lambda$       &           0.001          &  Outer update learning rate.     \\
			$E$             &             50           &  The overall training epoch.     \\
			$e$             &            100           &  The episodes in each epoch.     \\
			$B$             &             32           &  The batch size in each epoch.   \\
			$p$             &             25           &  \makecell[l]{``Patience index", the number of the episodes\\to wait after last time validation loss improved.}  \\
			$\mathcal{A}$   &             5            &  \makecell[l]{``Train-shot-attack-way", the number of attacks\\used in \textit{S} / \textit{S'}.} \\
			$\mathcal{A}_Q$ &             1            &  \makecell[l]{``Train-query-attack-way", the number of\\attacks used in \textit{Q} / \textit{Q'}.}       \\
			$k$             &            15            &  \makecell[l]{``Train-shot", the number of examples sampled\\per class in support set during training.}     \\
			$m$             &             6            &  \makecell[l]{``Train-query", the number of examples sampl-\\ed per class in query set during training.}      \\
			$K$             &             1            &  \makecell[l]{``Val/test-shot", the number of examples sam-\\pled per class in support set during testing or\\validating.}          \\
			$M$             &            15            &  \makecell[l]{``Val/test-query", the number of examples\\sampled per class in query set during testing\\or validating.}           \\
			\bottomrule[1pt]
		\end{tabular}
	\end{threeparttable}
\end{table}

\section{Experiment}

In this section, we conduct training of the models using the Meta-AT algorithm on the two MAD datasets. Subsequently, we compare the performance of Meta-AT with three SOTA defense methods. The experimental results indicate that Meta-AT surpasses the SOTA methods, achieving in the highest metric scores. Following the comparison experiments, we enhance our understanding by performing ablation studies to examine the impact of key parameter configurations in the algorithm.

\subsection{Setup}
\subsubsection*{\bf Equipment and basic parameter setting}
To assess the efficacy of Meta-AT, comprehensive experiments were carried out on the classification task. All experiments were performed on an HPC server equipped with an Nvidia Tesla P100 12G GPU and Intel (R) Xeon Gold 6132 CPU. The two datasets used for experimental validation are MAD-M and MAD-C, with ResNet-18 and AlexNet serving as the base classification backbone networks. The optimization process employed the SGD optimizer. For further reference, additional details regarding parameter settings can be found in Table~\ref{table_2}.

\subsubsection*{\bf Compared SOTA methods}
In the comparative experiments, we chose three different AT-based SOTA methods for comparison: the traditional AT, ATTA, and YOPO. ATTA mainly accumulates adversarial perturbations across epochs from previous epochs through a connection function $C$. With comparable model robustness, ATTA performed 12.2 times (14.1 times) faster than AT on MNIST (CIFAR10), and can also improve the accuracy of the AT-trained model against CIFAR10 by 7.2\%. YOPO effectively reduces the total number of fully forward and backward propagations by limiting the computation of adversarial samples to the first layer of the network. In about 20\% to 25\% GPU time, YOPO can achieve defense accuracy comparable to the PGD algorithm. The detailed parameter configurations of all comparison methods are consistent with those in the original source.

\subsection{Experimental Results Compared with SOTA Method}
\begin{figure}[t]
	\centering	
	\subfloat[]{\includegraphics[scale=0.3]{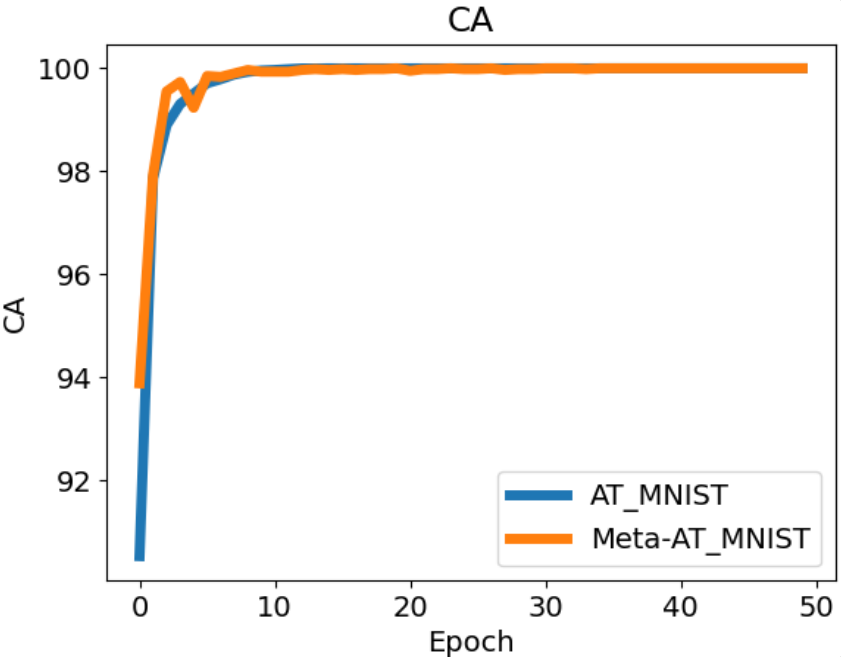}
		\label{fig_first_case}}
	\hfil
	\subfloat[]{\includegraphics[scale=0.3]{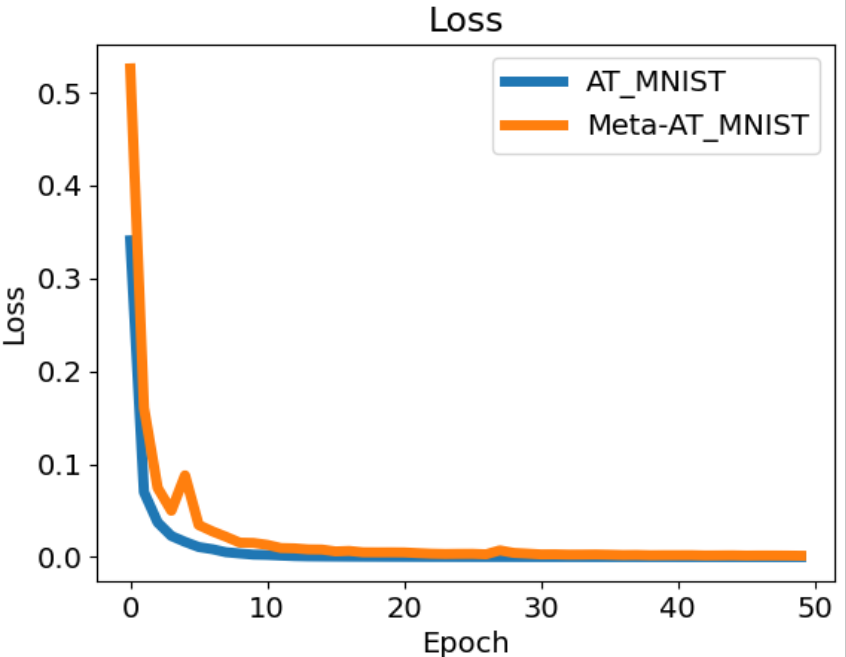}
		\label{fig_second_case}}
	\hfil
	\subfloat[]{\includegraphics[scale=0.3]{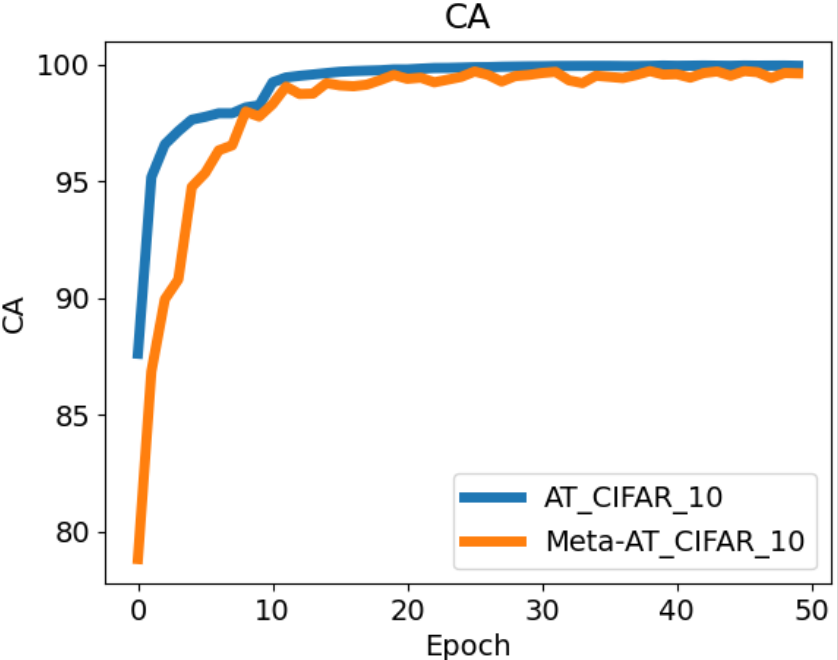}
		\label{fig_third_case}}
	\hfil
	\subfloat[]{\includegraphics[scale=0.3]{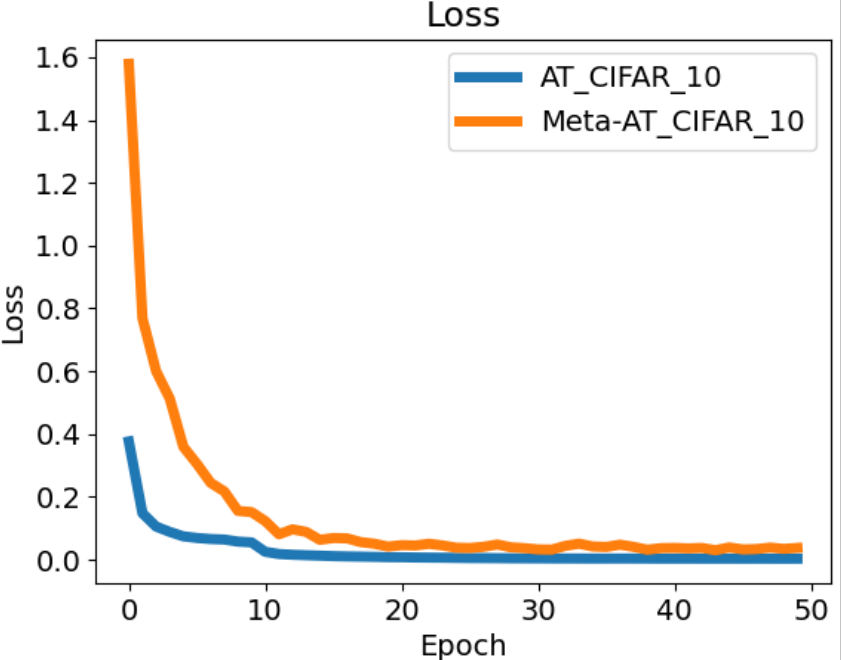}
		\label{fig_fourth_case}}
	\caption{The training loss and CA of Meta-AT and AT on two MAD datasets.}
	\label{fig_7}
\end{figure}
\begin{figure*}[t]
	\centering
	\subfloat[]{\includegraphics[scale=0.38]{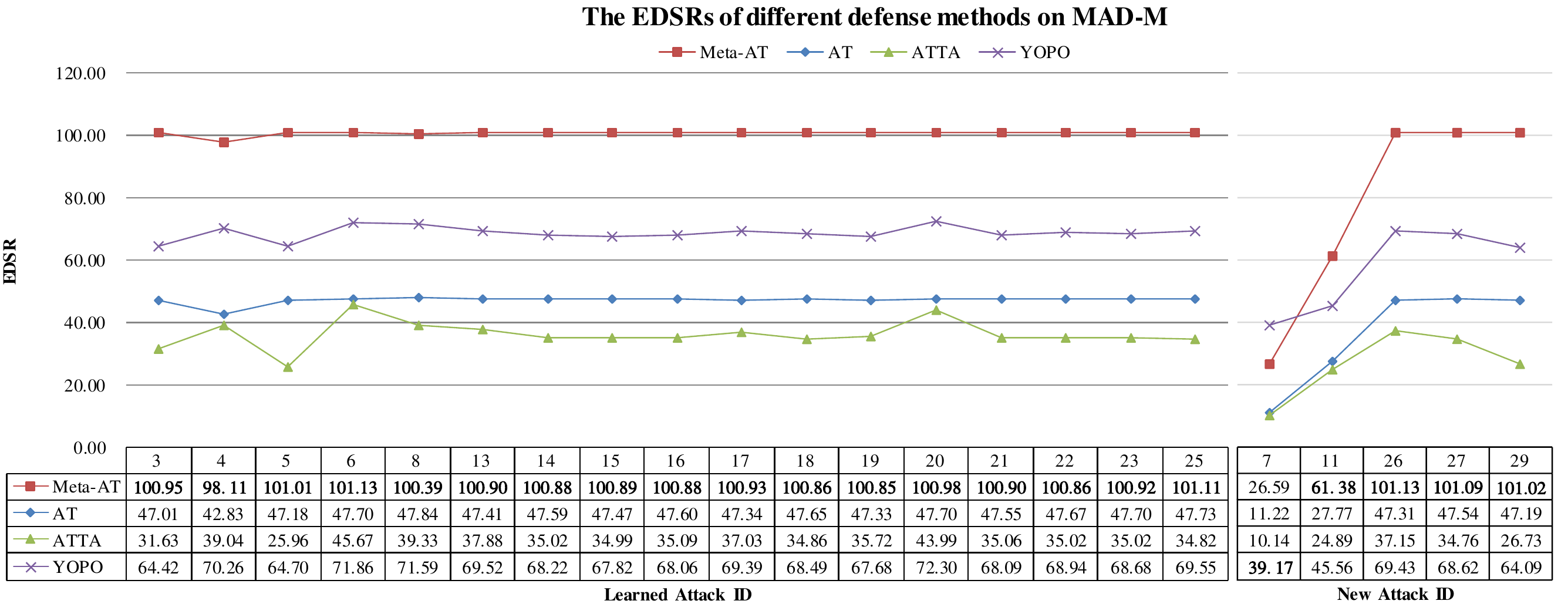}
		\label{fig_first_case}}
	\hfil
	\subfloat[]{\includegraphics[scale=0.48]{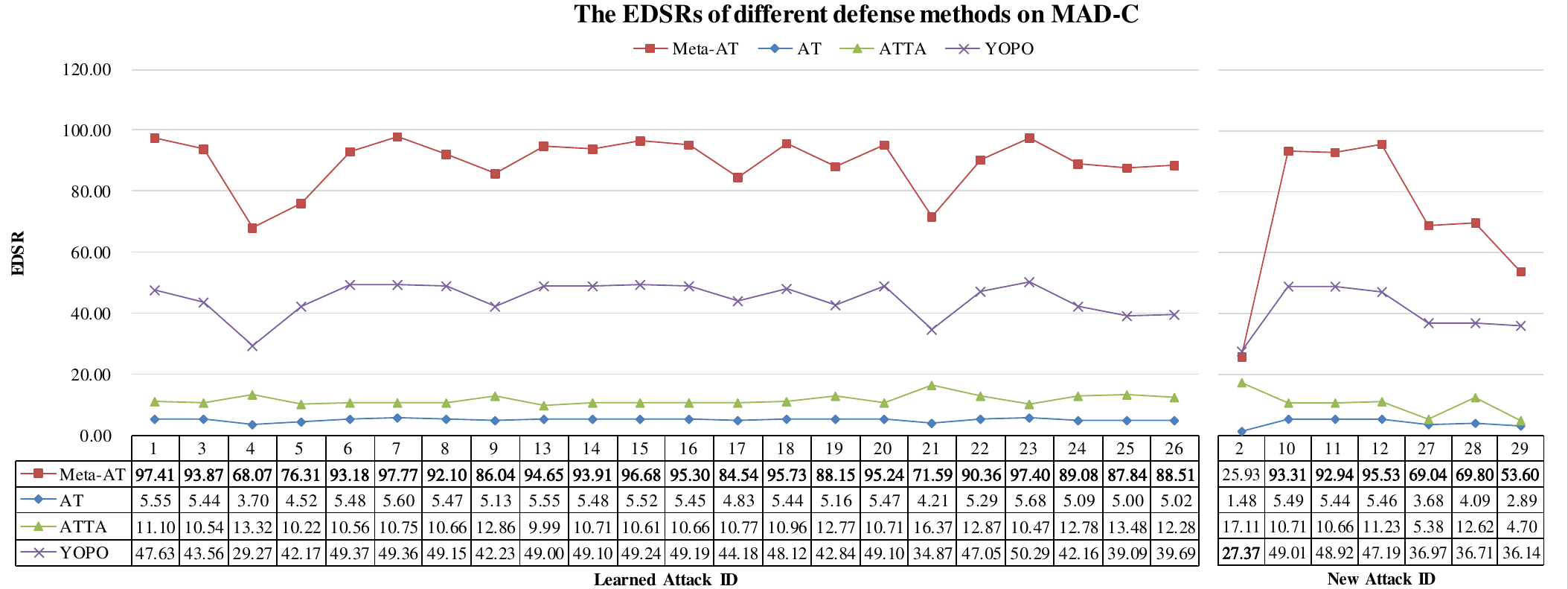}
		\label{fig_second_case}}
	\caption{The EDSRs of different defense methods on MAD datasets. (DSR exceeds 100\% because the models trained with Meta-AT have higher CA than the original CCA.)}
	\label{fig_8}
\end{figure*}
\subsubsection*{\bf Training progress}
The training processes of Meta-AT and AT on the two MAD datasets are shown in Fig.~\ref{fig_7}. In Meta-AT, the accuracy decreases significantly, and the loss increases noticeably during the learning of different attacks. However, when mini-ATs are applied within each meta-task, the model gradually adapts and learns to improve its performance, as indicated by the line fluctuations in the graphs. Furthermore, it is evident that the overall training process of the model gradually converges. These fluctuations and convergence align with our common sense and intuitively demonstrate the effectiveness of model learning. On the other hand, for AT, it is noticeable that the learning steps for adversarial examples in the case of AT are considerably larger compared to Meta-AT. Additionally, the rough and simple pattern used in AT results in a lower learning capability.

\subsubsection*{\bf Defense capability}
We initially assess different defense methods using the general DSR. Among them, AT employs the most straightforward fusion training strategy. ATTA and YOPO choose specific optimal combinations for two MAD datasets as mentioned in the original paper. ATTA adopts ``TRA40" for MAD-M and ``TRA10" for MAD-C, while YOPO adopts ``5-10" for MAD-M and ``5-3" for MAD-C. 

Table~\ref{table_3} presents the average DSR of various defense methods against both new and learned attacks. Notably, Meta-AT consistently demonstrates a high DSR across almost all attacks, surpassing even 100\% on MAD-M. This exceptional performance can be attributed to the fact that the model trained with Meta-AT possesses a higher CA compared to the original CCA. Overall, our Meta-AT secures the first rank, followed by AT, YOPO, and ATTA. (The optimal values are shown in bold.) To ensure a more comprehensive evaluation of the effectiveness of the defense methods, we have included the results of EDSR in Figure~\ref{fig_8}. (The optimal values are shown in bold.) The models trained with Meta-AT outperform all other SOTA methods. YOPO obtains the second rank but still lags behind Meta-AT by 50.31\% for MAD-C and 36.93\% for MAD-M in the worst cases. The best cases are achieved only on attack 7 for MAD-M and attack 2 for MAD-C. The defensive effects of ATTA and AT are relatively poorer. AT generally outperforms ATTA in MAD-M, while the opposite holds for MAD-C. The superior defensive effect of Meta-AT is more pronounced on MAD-M. This is mainly due to the more complex data distribution of CIFAR-10 compared to MNIST, which renders it more susceptible to adversarial attacks and subsequently more challenging to enhance the defenses against them. In summary, the models trained with Meta-AT show excellent defensive effects against both learned and new attacks.
\begin{table}
	\caption{The average DSRs (\%) of different defense methods on MAD datasets.\label{table_3}}
	\begin{threeparttable}
		\begin{tabular*}{\linewidth}{@{\extracolsep\fill}ccccc}	
			\toprule[1pt] 
			\multicolumn{1}{l}{\multirow{2}{*}{\textbf{\makecell[c]{Defence \\ Method }}}}&\multicolumn{2}{@{}c@{}}{\textbf{MAD-M}}&\multicolumn{2}{@{}c@{}}{\textbf{MAD-C}}        \\
			\cmidrule{2-3}\cmidrule{4-5} 
			~ & Learned  & New  & Learned  & New   \\
			\midrule [0.5pt] 
			Meta-AT & \textbf{101.4}  & \textbf{79.6} & \textbf{90.7} & \textbf{73.0}  \\
			AT      & 98.4            & 75.4          & 79.8          & 62.7           \\ 
			ATTA    & 36.8            & 27.1          & 11.7          & 10.4           \\ 
			YOPO    & 84.9            & 72.8          & 52.7          & 47.4           \\                 
			\bottomrule[1pt] 
		\end{tabular*}
			
		\begin{tablenotes}
			\footnotesize
			\item DSR exceeds 100\% because the models trained with Meta-AT have higher CA than the original CCA.
		\end{tablenotes}
	\end{threeparttable}
\end{table}
\subsubsection*{\bf Clean-samples classification accuracy}
If a model needs to sacrifice the accuracy of clean examples to achieve a high defensive capability, we consider it to be not robust enough. Another key strength of Meta-AT is that it maintains a high CCA. Table \ref{table_4} provides the CCA of the model trained before and after applying various defense methods on original test datasets. (The optimal values are shown in bold.) For Meta-AT, we provide the range of CCA achieved when different attacks are applied. As observed from the table, all the defense methods lead to a decrease in the CCA of the original model. However, Meta-AT has minimal impact on the original model and maintains a stable high CCA across different attacks, without overfitting to a specific attack. It exhibits high robustness compared to models trained with other SOTA methods. 

\subsubsection*{\bf Operating time}
In the current deep learning environment, saving training costs in terms of computational resources is crucial. The operating time of Meta-AT is primarily impacted by the number of episodes, as evident from the testing process of Meta-AT for a single attack shown in Figure \ref{fig_9}. The learning of 60 steps for different attacks tends to stabilize, taking about 2 seconds per step. The early stopping mechanism has also improved the optimization of the learning period in this regard. The training times of Meta-AT and other SOTA defense methods are presented in Table \ref{table_4}. (The optimal values are shown in bold.) With Meta-AT, learning a few adversarial examples per class requires approximately 0.33 minutes for MAD-M and 2 minutes for MAD-C, allowing the model to achieve the highest CA. ATTA is the closest method in terms of training time, slightly faster than Meta-AT in some cases, followed by YOPO, while AT is the slowest. Overall, Meta-AT significantly shortens the learning period, and combined with its high DSR and CCA establishes itself as a leading method in industrial-grade AT-based defense methods.
\begin{table}
	\caption{The average CCAs (\%) and OTs (h) of different defense methods on MAD datasets.\label{table_4}}
	\begin{tabular*}{\linewidth}{@{\extracolsep\fill}ccccc}	
		\toprule[1pt] 
		\multicolumn{1}{l}{\multirow{2}{*}{\textbf{\makecell[c]{Defence \\ Method }}}}&\multicolumn{2}{@{}c@{}}{\textbf{MAD-M}}&\multicolumn{2}{@{}c@{}}{\textbf{MAD-C}}        \\
		\cmidrule(r){2-3}\cmidrule(l){4-5}
		          ~ & CCA    & OT     & CCA    & OT   \\
		\midrule [0.5pt] 
		No-defense  & 98.84 & -      & 94.82 & -    \\
		AT          & 88.20 & 0.7333 & 82.75 & 2.7333       \\ 
		ATTA        & 35.67 & 0.0143 & 10.20 & \textbf{0.0090}  \\ 
		YOPO        & 87.93 & 0.2383 & 55.25 & 0.1612 \\   
		\textbf{Meta-AT} & \textbf{96.06$\pm$0.08} & \textbf{0.0055} & \textbf{88.60$\pm$0.05} & 0.0210  \\              
		\bottomrule[1pt] 
	\end{tabular*}
\end{table}
\subsection{Ablation Study}
To optimize the benefits of the Meta-AT model, we conducted ablation studies on the following four key aspects. 

\subsubsection*{\bf Learning rate}
We first determine the effective combination of inner and outer update learning rates, $\beta$ and $\lambda$. When using $(\beta=0.01, \lambda=0.01)$ as the learning rates, the inner update did not converge. The outer update was slower compared to the first combination when $(\beta=0.001, \lambda=0.001)$ was used. Aligns with experimental experience, which suggests that utilizing larger steps for learning optimization in the outer update and smaller steps for fine-tuning in the inner update is a beneficial approach. Therefore, the combination of $(\beta=0.01, \lambda=0.001)$ led to an effective convergence of the model.
\begin{figure}
	\centering	
	\includegraphics[scale=0.295]{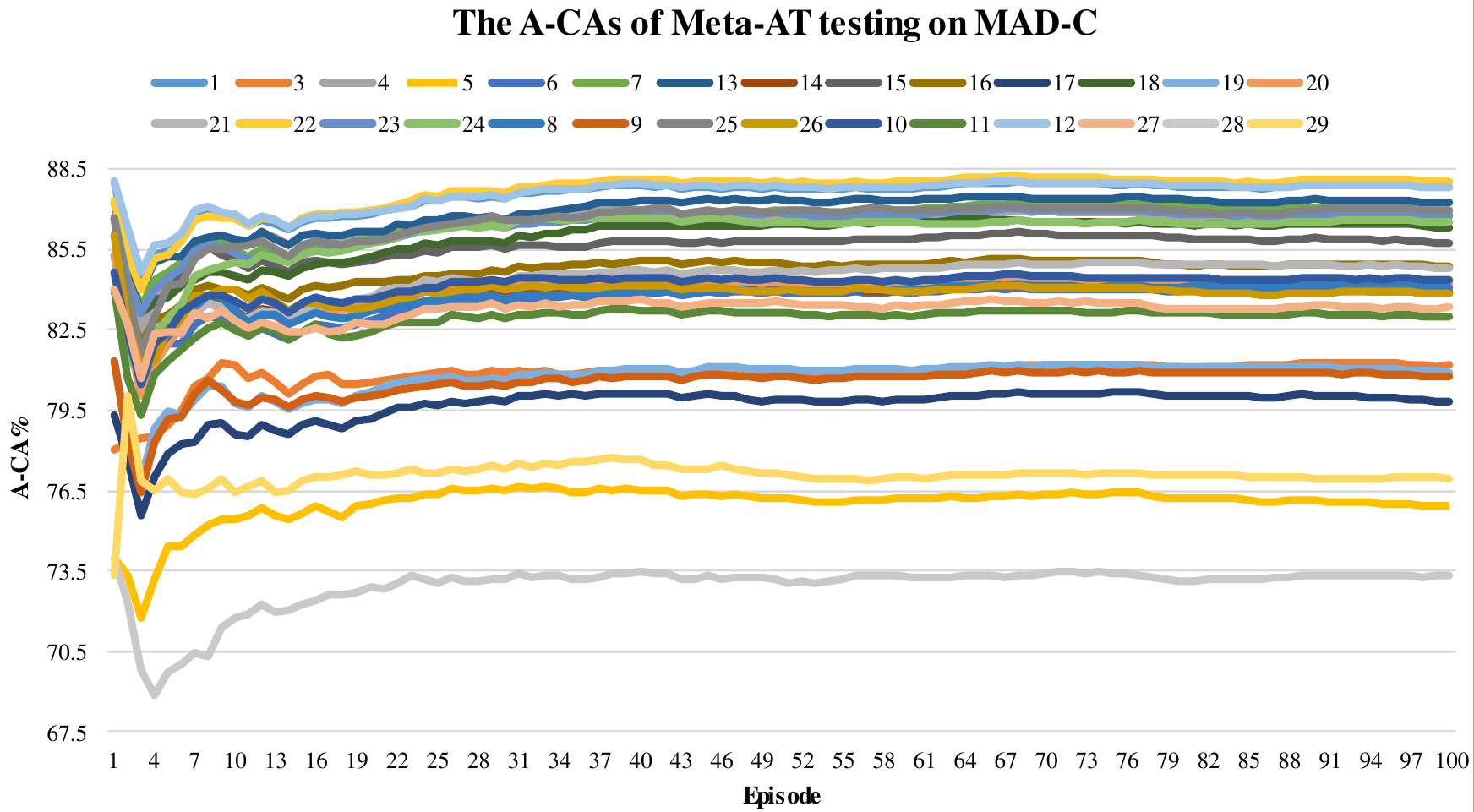}
	\caption{The A-CAs of the \textit{``5-way, 1-shot"} Meta-AT testing on the test examples of MAD-M and MAD-C.}
	\label{fig_9}
\end{figure}
\subsubsection*{\bf Patience index ($p$)}
The testing process of 100 episodes on the test examples of MAD-M and MAD-C are shown in Fig.~\ref{fig_9}. The graph reveals significant fluctuations during the initial learning stage. The overall trend of learning is characterized by an initial decrease, followed by an increase, and then another decrease. The initial decrease to increase reflects the pattern of adversarial training, while the subsequent decrease indicates the gradual overfitting of the model. To prevent model overfitting and improve EDSR, an appropriate value of \textit{p} in the early stopping mechanism is crucial. Based on experimental experience, we set $p=25$ to ensure the effectiveness of Meta-AT.
\subsubsection*{\bf \textit{``$\mathcal{A}$-way, K-shot"}}
\begin{figure*}[t]
	\centering
	\includegraphics[scale=0.5]{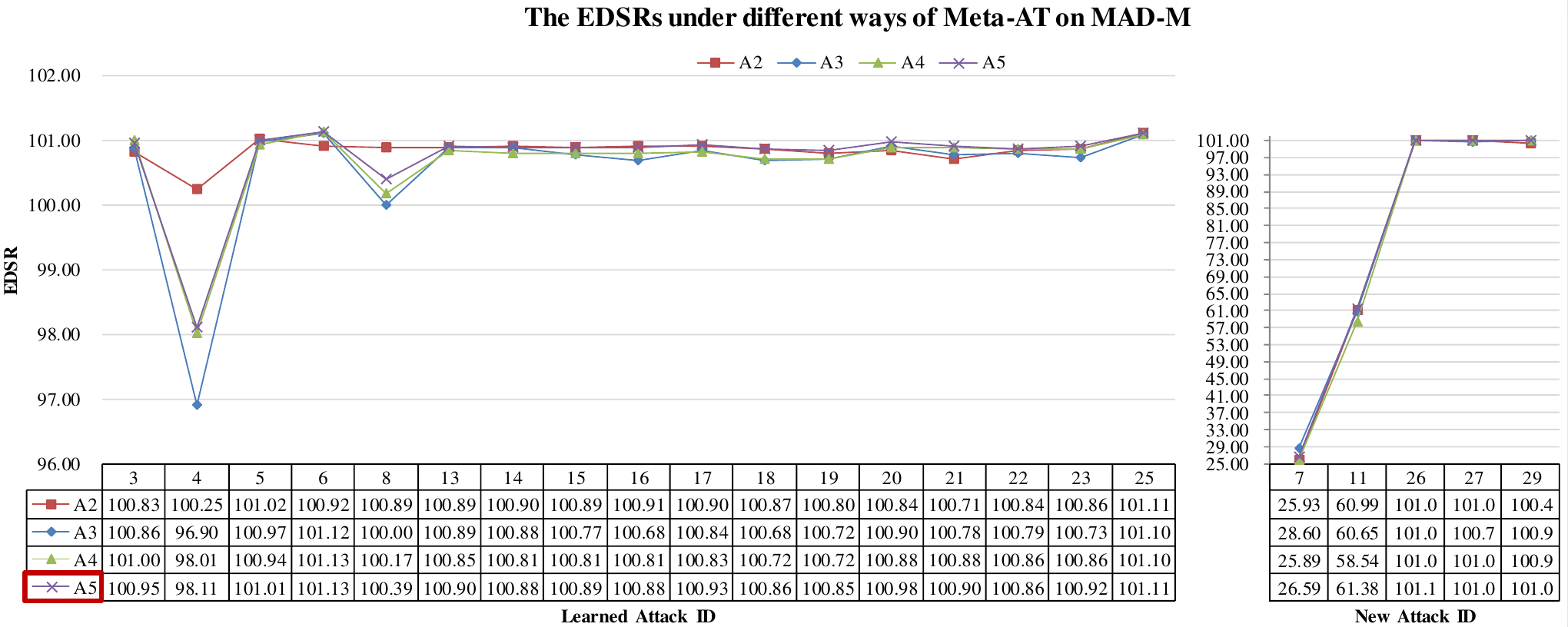}
	\caption{The EDSRs under different ways of Meta-AT on MAD-M.}
	\label{fig_10}
\end{figure*}

This paper mainly studies a learnable model under the \textit{``$\mathcal{A}$-way,  K-shot"} mode by using adversarial samples of $\mathcal{A}$ known attacks. New attacks can achieve a higher defense rate by fine-tuning this model with \textit{K} samples. Among them, the value of \textit{K} is derived from the context of few-shot classification tasks. Hence, we need to study the setting of parameter $\mathcal{A}$. The selection of $\mathcal{A}$ as 2, 3, 4, and 5 is based on MAD-M, excluding $\mathcal{A}=1$ due to its similarity to the traditional AT mode. Fig.~\ref{fig_10} displays the EDSRs for different Meta-AT models ($\mathcal{A}=2,3,4,5$ correspond to A2, A3, A4, and A5 in the figure). It can be observed that the impact of $\mathcal{A}$ on model training is limited, with notable fluctuations occurring in only a few Meta-ATs for known attacks (e.g., attacks 4 and 8) and less impact on new attacks. Fig.~\ref{fig_10} alone is insufficient to determine a suitable $\mathcal{A}$ value. Therefore, we comprehensively consider the CCA across different models and calculate the average EDSRs to differentiate them, as presented in Table~\ref{table_5}. From Table~\ref{table_5}, when $\mathcal{A}=5$, there are the highest CCA and EDSR are observed (indicated in bold). In conclusion, the \textit{``5-way, 1-shot"} pattern serves as the basis for the following comparative experiments.

\begin{table}[h]
	\caption{The CCAs and average EDSRs under different ways of Meta-AT in MAD-M.\label{table_5}}
	\centering
	\begin{threeparttable}
		\begin{tabular}{lll}
			\toprule[1pt]  
					\textbf{Way}     &         \textbf{CCA}         &     \textbf{EDSR}     \\
			\midrule[0.5pt]	
			A2                 &               0.9950          &            89.37        \\
			A3                 &               0.9955          &            89.44        \\
			A4                 &               0.9965          &            89.08        \\
			A5                 &        \textbf{0.9965}        &     \textbf{89.49}      \\
			\bottomrule[1pt]	
		\end{tabular}
	\end{threeparttable}
\end{table}

\begin{figure*}[t]
	\centering
	\includegraphics[scale=0.51]{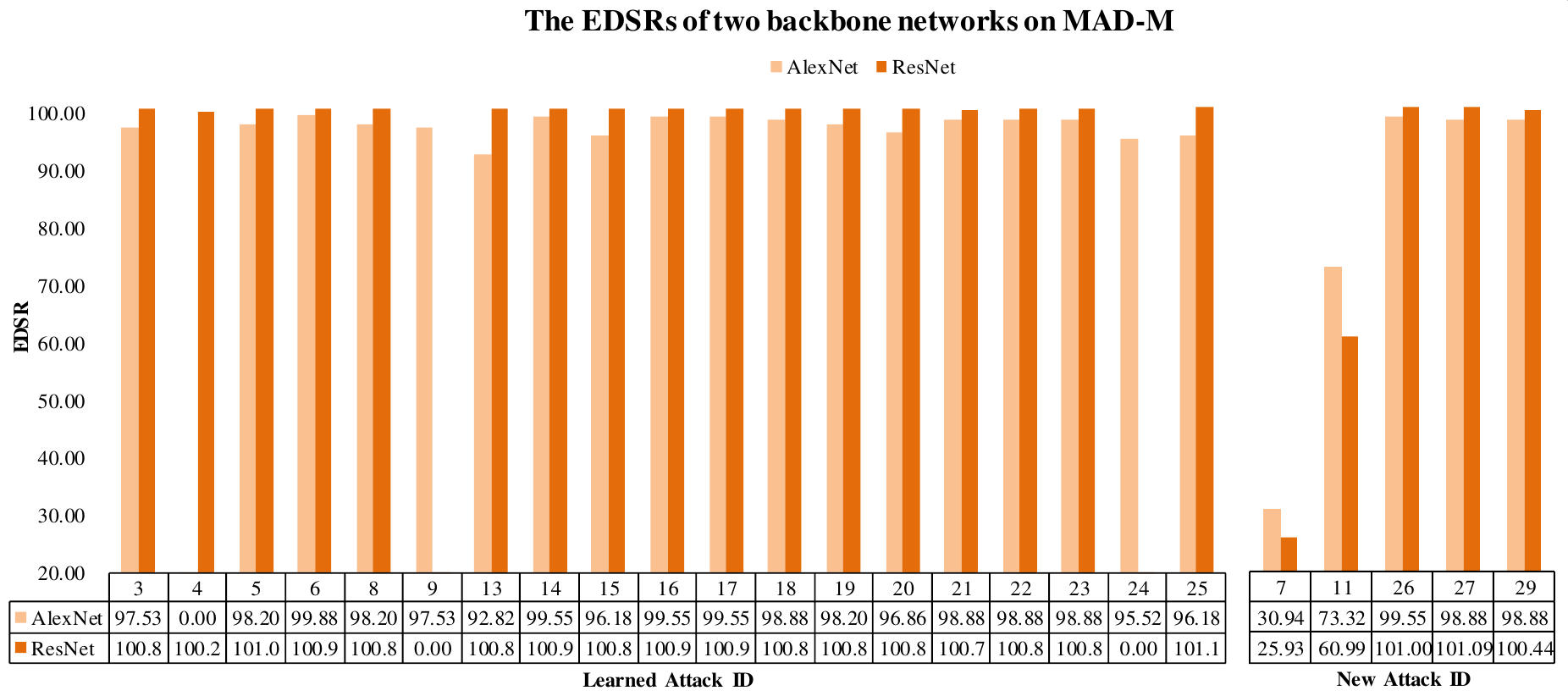}
	\caption{The EDSRs under different backbone networks of Meta-AT on MAD-M.}
	\label{fig_11}
\end{figure*}
\subsubsection*{\bf Classification backbone network}
To assess the effectiveness of our proposed Meta-AT on different classification backbone networks, we also created the MAD-M dataset based on AlexNet. Using the aforementioned parameters, Fig.~\ref{fig_11} shows the EDSRs of different attacks after applying adversarial examples fine-tuned on both AlexNet and ResNet-18 under the Meta-AT in a \textit{``5-way, 1-shot"} setting. The graph indicates that attack 4 is ineffective on AlexNet, whereas attacks 9 and 24 are ineffective on ResNet-18. This discrepancy can be attributed to the relatively simple distribution of the MNIST dataset, which makes the simpler AlexNet network better suited for this task and most attacks effective against it, while encountering fewer ineffective attacks. In terms of EDSR, the ResNet-18-based EDSR consistently outperforms that of AlexNet. This is attributed to the higher complexity and larger ``capacity" of ResNet-18, resulting in higher CA. For new attacks, attacks 7 and 11 exhibit strong adversarial characteristics, and AlexNet demonstrates superior adversarial defense capabilities compared to ResNet-18. Overall, these experiments demonstrate the transferability of adversarial defense through Meta-AT across different classification backbone networks. In the previous SOTA comparison experiments, we standardized the backbone network to the higher capacity ResNet-18 to control variables.

\section{Conclusion}

This paper introduces a novel MAD benchmark consisting of two extensive MAD datasets, a MAD evaluation protocol, and a baseline algorithm called Meta-AT. The MAD datasets are created by subjecting the MNIST and CIFAR-10 datasets to 30 mainstream adversarial attacks. In the evaluation protocol, besides introducing some dataset configurations, we have also innovatively proposed a versatile evaluation metric called EDSR, which provides a comprehensive and unbiased assessment of the effectiveness of defense methods. Meta-AT exhibits the capability of ``predictive defense". It investigates the transferability of adversarial defense methods to new attacks and the ability to learn from a limited number of adversarial examples. Experimental results demonstrate that the Meta-AT model achieves a significantly highest EDSR compared with other SOTA methods against both known and unknown attacks through few-shot learning within a few minutes, while maintaining a stable and high CCA across various types of attacks. These performance levels are comparable to those of industrial-grade model defense applications, which is a significant accomplishment. However, one limitation is that the MADs created for Meta-AT are quite large, and future work will focus on reducing their size while maintaining comparable learning effectiveness. Furthermore, we also plan to optimize the Meta-AT framework and explore its application to other tasks such as object detection and semantic segmentation.

\bibliography{mybibfile}

\begin{thebibliography}{10}

\bibitem{wang2022review}
Z.~Wang, J.~Zhan, C.~Duan, X.~Guan, P.~Lu, and K.~Yang, ``A review of vehicle
  detection techniques for intelligent vehicles,'' {\em IEEE Transactions on
  Neural Networks and Learning Systems}, 2022.

\bibitem{bouchaffra2014nonlinear}
D.~Bouchaffra, ``Nonlinear topological component analysis: application to
  age-invariant face recognition,'' {\em IEEE transactions on neural networks
  and learning systems}, vol.~26, no.~7, pp.~1375--1387, 2014.

\bibitem{zhou2020human}
T.~Zhou, J.~Shen, D.~He, P.~Vijayakumar, and N.~Kumar,
  ``Human-in-the-loop-aided privacy-preserving scheme for smart healthcare,''
  {\em IEEE transactions on emerging topics in computational intelligence},
  vol.~6, no.~1, pp.~6--15, 2020.

\bibitem{akhtar2021advances}
N.~Akhtar, A.~Mian, N.~Kardan, and M.~Shah, ``Advances in adversarial attacks
  and defenses in computer vision: A survey,'' {\em IEEE Access}, vol.~9,
  pp.~155161--155196, 2021.

\bibitem{tao2018attacks}
G.~Tao, S.~Ma, Y.~Liu, and X.~Zhang, ``Attacks meet interpretability:
  Attribute-steered detection of adversarial samples,'' {\em Advances in Neural
  Information Processing Systems}, vol.~31, 2018.

\bibitem{li2020connecting}
S.~Li, S.~Zhu, S.~Paul, A.~Roy-Chowdhury, C.~Song, S.~Krishnamurthy, A.~Swami,
  and K.~S. Chan, ``Connecting the dots: Detecting adversarial perturbations
  using context inconsistency,'' in {\em Computer Vision--ECCV 2020: 16th
  European Conference, Glasgow, UK, August 23--28, 2020, Proceedings, Part
  XXIII 16}, pp.~396--413, Springer, 2020.

\bibitem{guo2017countering}
C.~Guo, M.~Rana, M.~Cisse, and L.~Van Der~Maaten, ``Countering adversarial
  images using input transformations,'' {\em arXiv preprint arXiv:1711.00117},
  2017.

\bibitem{raff2019barrage}
E.~Raff, J.~Sylvester, S.~Forsyth, and M.~McLean, ``Barrage of random
  transforms for adversarially robust defense,'' in {\em Proceedings of the
  IEEE/CVF Conference on Computer Vision and Pattern Recognition},
  pp.~6528--6537, 2019.

\bibitem{zhai2020macer}
R.~Zhai, C.~Dan, D.~He, H.~Zhang, B.~Gong, P.~Ravikumar, C.-J. Hsieh, and
  L.~Wang, ``Macer: Attack-free and scalable robust training via maximizing
  certified radius,'' {\em arXiv preprint arXiv:2001.02378}, 2020.

\bibitem{jia2019certified}
J.~Jia, X.~Cao, B.~Wang, and N.~Z. Gong, ``Certified robustness for top-k
  predictions against adversarial perturbations via randomized smoothing,''
  {\em arXiv preprint arXiv:1912.09899}, 2019.

\bibitem{cemgil2019adversarially}
T.~Cemgil, S.~Ghaisas, K.~D. Dvijotham, and P.~Kohli, ``Adversarially robust
  representations with smooth encoders,'' in {\em International Conference on
  Learning Representations}, 2019.

\bibitem{he2020defending}
Z.~He, A.~S. Rakin, J.~Li, C.~Chakrabarti, and D.~Fan, ``Defending and
  harnessing the bit-flip based adversarial weight attack,'' in {\em
  Proceedings of the IEEE/CVF Conference on Computer Vision and Pattern
  Recognition}, pp.~14095--14103, 2020.

\bibitem{madry2017towards}
A.~Madry, A.~Makelov, L.~Schmidt, D.~Tsipras, and A.~Vladu, ``Towards deep
  learning models resistant to adversarial attacks,'' {\em arXiv preprint
  arXiv:1706.06083}, 2017.

\bibitem{shafahi2019adversarial}
A.~Shafahi, M.~Najibi, M.~A. Ghiasi, Z.~Xu, J.~Dickerson, C.~Studer, L.~S.
  Davis, G.~Taylor, and T.~Goldstein, ``Adversarial training for free!,'' {\em
  Advances in Neural Information Processing Systems}, vol.~32, 2019.

\bibitem{maini2020adversarial}
P.~Maini, E.~Wong, and Z.~Kolter, ``Adversarial robustness against the union of
  multiple perturbation models,'' in {\em International Conference on Machine
  Learning}, pp.~6640--6650, PMLR, 2020.

\bibitem{farnia2018generalizable}
F.~Farnia, J.~M. Zhang, and D.~Tse, ``Generalizable adversarial training via
  spectral normalization,'' {\em arXiv preprint arXiv:1811.07457}, 2018.

\bibitem{lecun1998gradient}
Y.~LeCun, L.~Bottou, Y.~Bengio, and P.~Haffner, ``Gradient-based learning
  applied to document recognition,'' {\em Proceedings of the IEEE}, vol.~86,
  no.~11, pp.~2278--2324, 1998.

\bibitem{krizhevsky2009learning}
A.~Krizhevsky, G.~Hinton, {\em et~al.}, ``Learning multiple layers of features
  from tiny images,'' 2009.

\bibitem{zheng2020efficient}
H.~Zheng, Z.~Zhang, J.~Gu, H.~Lee, and A.~Prakash, ``Efficient adversarial
  training with transferable adversarial examples,'' in {\em Proceedings of the
  IEEE/CVF Conference on Computer Vision and Pattern Recognition},
  pp.~1181--1190, 2020.

\bibitem{zhang2019you}
D.~Zhang, T.~Zhang, Y.~Lu, Z.~Zhu, and B.~Dong, ``You only propagate once:
  Accelerating adversarial training via maximal principle,'' {\em Advances in
  Neural Information Processing Systems}, vol.~32, 2019.

\bibitem{miyato2018virtual}
T.~Miyato, S.-i. Maeda, M.~Koyama, and S.~Ishii, ``Virtual adversarial
  training: a regularization method for supervised and semi-supervised
  learning,'' {\em IEEE transactions on pattern analysis and machine
  intelligence}, vol.~41, no.~8, pp.~1979--1993, 2018.

\bibitem{meng2023integrating}
J.~Meng, F.~Zhu, Y.~Ge, and P.~Zhao, ``Integrating safety constraints into
  adversarial training for robust deep reinforcement learning,'' {\em
  Information Sciences}, vol.~619, pp.~310--323, 2023.

\bibitem{ke2019araml}
P.~Ke, F.~Huang, M.~Huang, and X.~Zhu, ``Araml: A stable adversarial training
  framework for text generation,'' {\em arXiv preprint arXiv:1908.07195}, 2019.

\bibitem{wang2021visual}
T.~Wang, Z.~Wu, and D.~Wang, ``Visual perception generalization for
  vision-and-language navigation via meta-learning,'' {\em IEEE Transactions on
  Neural Networks and Learning Systems}, 2021.

\bibitem{yu2023pid}
Z.~Yu and G.~Sun, ``A pid based meta-learning method about space
  non-cooperative active object tracking,'' {\em IEEE Transactions on Vehicular
  Technology}, 2023.

\bibitem{goldblum2020adversarially}
M.~Goldblum, L.~Fowl, and T.~Goldstein, ``Adversarially robust few-shot
  learning: A meta-learning approach,'' {\em Advances in Neural Information
  Processing Systems}, vol.~33, pp.~17886--17895, 2020.

\bibitem{liu2021long}
F.~Liu, S.~Zhao, X.~Dai, and B.~Xiao, ``Long-term cross adversarial training: A
  robust meta-learning method for few-shot classification tasks,'' {\em arXiv
  preprint arXiv:2106.12900}, 2021.

\bibitem{qi2022cross}
J.~Qi, R.~Zhang, C.~Li, and Y.~Mao, ``Cross domain few-shot learning via meta
  adversarial training,'' {\em arXiv preprint arXiv:2202.05713}, 2022.

\bibitem{ma2019metaadvdet}
C.~Ma, C.~Zhao, H.~Shi, L.~Chen, J.~Yong, and D.~Zeng, ``Metaadvdet: Towards
  robust detection of evolving adversarial attacks,'' in {\em Proceedings of
  the 27th ACM International Conference on Multimedia}, pp.~692--701, 2019.

\bibitem{metzen2021meta}
J.~H. Metzen, N.~Finnie, and R.~Hutmacher, ``Meta adversarial training against
  universal patches,'' {\em arXiv preprint arXiv:2101.11453}, 2021.

\bibitem{papernot2016limitations}
N.~Papernot, P.~McDaniel, S.~Jha, M.~Fredrikson, Z.~B. Celik, and A.~Swami,
  ``The limitations of deep learning in adversarial settings,'' in {\em 2016
  IEEE European symposium on security and privacy (EuroS\&P)}, pp.~372--387,
  IEEE, 2016.

\bibitem{moosavi2016deepfool}
S.-M. Moosavi-Dezfooli, A.~Fawzi, and P.~Frossard, ``Deepfool: a simple and
  accurate method to fool deep neural networks,'' in {\em Proceedings of the
  IEEE conference on computer vision and pattern recognition}, pp.~2574--2582,
  2016.

\bibitem{moosavi2017universal}
S.-M. Moosavi-Dezfooli, A.~Fawzi, O.~Fawzi, and P.~Frossard, ``Universal
  adversarial perturbations,'' in {\em Proceedings of the IEEE conference on
  computer vision and pattern recognition}, pp.~1765--1773, 2017.

\bibitem{jang2017objective}
U.~Jang, X.~Wu, and S.~Jha, ``Objective metrics and gradient descent algorithms
  for adversarial examples in machine learning,'' in {\em Proceedings of the
  33rd Annual Computer Security Applications Conference}, pp.~262--277, 2017.

\bibitem{brendel2017decision}
W.~Brendel, J.~Rauber, and M.~Bethge, ``Decision-based adversarial attacks:
  Reliable attacks against black-box machine learning models,'' {\em arXiv
  preprint arXiv:1712.04248}, 2017.

\bibitem{chen2018ead}
P.-Y. Chen, Y.~Sharma, H.~Zhang, J.~Yi, and C.-J. Hsieh, ``Ead: elastic-net
  attacks to deep neural networks via adversarial examples,'' in {\em
  Proceedings of the AAAI conference on artificial intelligence}, vol.~32,
  2018.

\bibitem{chen2017zoo}
P.-Y. Chen, H.~Zhang, Y.~Sharma, J.~Yi, and C.-J. Hsieh, ``Zoo: Zeroth order
  optimization based black-box attacks to deep neural networks without training
  substitute models,'' in {\em Proceedings of the 10th ACM workshop on
  artificial intelligence and security}, pp.~15--26, 2017.

\bibitem{engstrom2019exploring}
L.~Engstrom, B.~Tran, D.~Tsipras, L.~Schmidt, and A.~Madry, ``Exploring the
  landscape of spatial robustness,'' in {\em International conference on
  machine learning}, pp.~1802--1811, PMLR, 2019.

\bibitem{chen2020hopskipjumpattack}
J.~Chen, M.~I. Jordan, and M.~J. Wainwright, ``Hopskipjumpattack: A
  query-efficient decision-based attack,'' in {\em 2020 ieee symposium on
  security and privacy (sp)}, pp.~1277--1294, IEEE, 2020.

\bibitem{guo2019simple}
C.~Guo, J.~Gardner, Y.~You, A.~G. Wilson, and K.~Weinberger, ``Simple black-box
  adversarial attacks,'' in {\em International Conference on Machine Learning},
  pp.~2484--2493, PMLR, 2019.

\bibitem{ghiasi2020breaking}
A.~Ghiasi, A.~Shafahi, and T.~Goldstein, ``Breaking certified defenses:
  Semantic adversarial examples with spoofed robustness certificates,'' {\em
  arXiv preprint arXiv:2003.08937}, 2020.

\bibitem{rahmati2020geoda}
A.~Rahmati, S.-M. Moosavi-Dezfooli, P.~Frossard, and H.~Dai, ``Geoda: a
  geometric framework for black-box adversarial attacks,'' in {\em Proceedings
  of the IEEE/CVF conference on computer vision and pattern recognition},
  pp.~8446--8455, 2020.

\bibitem{wong2019wasserstein}
E.~Wong, F.~Schmidt, and Z.~Kolter, ``Wasserstein adversarial examples via
  projected sinkhorn iterations,'' in {\em International Conference on Machine
  Learning}, pp.~6808--6817, PMLR, 2019.

\bibitem{goodfellow2014explaining}
I.~J. Goodfellow, J.~Shlens, and C.~Szegedy, ``Explaining and harnessing
  adversarial examples,'' {\em arXiv preprint arXiv:1412.6572}, 2014.

\bibitem{kurakin2018adversarial}
A.~Kurakin, I.~J. Goodfellow, and S.~Bengio, ``Adversarial examples in the
  physical world,'' in {\em Artificial intelligence safety and security},
  pp.~99--112, Chapman and Hall/CRC, 2018.

\bibitem{carlini2017towards}
N.~Carlini and D.~Wagner, ``Towards evaluating the robustness of neural
  networks,'' in {\em 2017 ieee symposium on security and privacy (sp)},
  pp.~39--57, Ieee, 2017.

\bibitem{dong2018boosting}
Y.~Dong, F.~Liao, T.~Pang, H.~Su, J.~Zhu, X.~Hu, and J.~Li, ``Boosting
  adversarial attacks with momentum,'' in {\em Proceedings of the IEEE
  conference on computer vision and pattern recognition}, pp.~9185--9193, 2018.

\bibitem{dong2019evading}
Y.~Dong, T.~Pang, H.~Su, and J.~Zhu, ``Evading defenses to transferable
  adversarial examples by translation-invariant attacks,'' in {\em Proceedings
  of the IEEE/CVF Conference on Computer Vision and Pattern Recognition},
  pp.~4312--4321, 2019.

\bibitem{zhang2019theoretically}
H.~Zhang, Y.~Yu, J.~Jiao, E.~Xing, L.~El~Ghaoui, and M.~Jordan, ``Theoretically
  principled trade-off between robustness and accuracy,'' in {\em International
  conference on machine learning}, pp.~7472--7482, PMLR, 2019.

\bibitem{tramer2017ensemble}
F.~Tram{\`e}r, A.~Kurakin, N.~Papernot, I.~Goodfellow, D.~Boneh, and
  P.~McDaniel, ``Ensemble adversarial training: Attacks and defenses,'' {\em
  arXiv preprint arXiv:1705.07204}, 2017.

\bibitem{croce2020reliable}
F.~Croce and M.~Hein, ``Reliable evaluation of adversarial robustness with an
  ensemble of diverse parameter-free attacks,'' in {\em International
  conference on machine learning}, pp.~2206--2216, PMLR, 2020.

\bibitem{wong2020fast}
E.~Wong, L.~Rice, and J.~Z. Kolter, ``Fast is better than free: Revisiting
  adversarial training,'' {\em arXiv preprint arXiv:2001.03994}, 2020.

\bibitem{andriushchenko2020square}
M.~Andriushchenko, F.~Croce, N.~Flammarion, and M.~Hein, ``Square attack: a
  query-efficient black-box adversarial attack via random search,'' in {\em
  Computer Vision--ECCV 2020: 16th European Conference, Glasgow, UK, August
  23--28, 2020, Proceedings, Part XXIII}, pp.~484--501, Springer, 2020.

\bibitem{liu2018adv}
X.~Liu, Y.~Li, C.~Wu, and C.-J. Hsieh, ``Adv-bnn: Improved adversarial defense
  through robust bayesian neural network,'' {\em arXiv preprint
  arXiv:1810.01279}, 2018.

\bibitem{su2019one}
J.~Su, D.~V. Vargas, and K.~Sakurai, ``One pixel attack for fooling deep neural
  networks,'' {\em IEEE Transactions on Evolutionary Computation}, vol.~23,
  no.~5, pp.~828--841, 2019.

\bibitem{croce2020minimally}
F.~Croce and M.~Hein, ``Minimally distorted adversarial examples with a fast
  adaptive boundary attack,'' in {\em International Conference on Machine
  Learning}, pp.~2196--2205, PMLR, 2020.

\bibitem{rusu2016progressive}
A.~A. Rusu, N.~C. Rabinowitz, G.~Desjardins, H.~Soyer, J.~Kirkpatrick,
  K.~Kavukcuoglu, R.~Pascanu, and R.~Hadsell, ``Progressive neural networks,''
  {\em arXiv preprint arXiv:1606.04671}, 2016.

\end{thebibliography}

\vspace{-20pt}
\begin{IEEEbiography}
[{\includegraphics[width=1in,height=1.25in,clip,keepaspectratio]{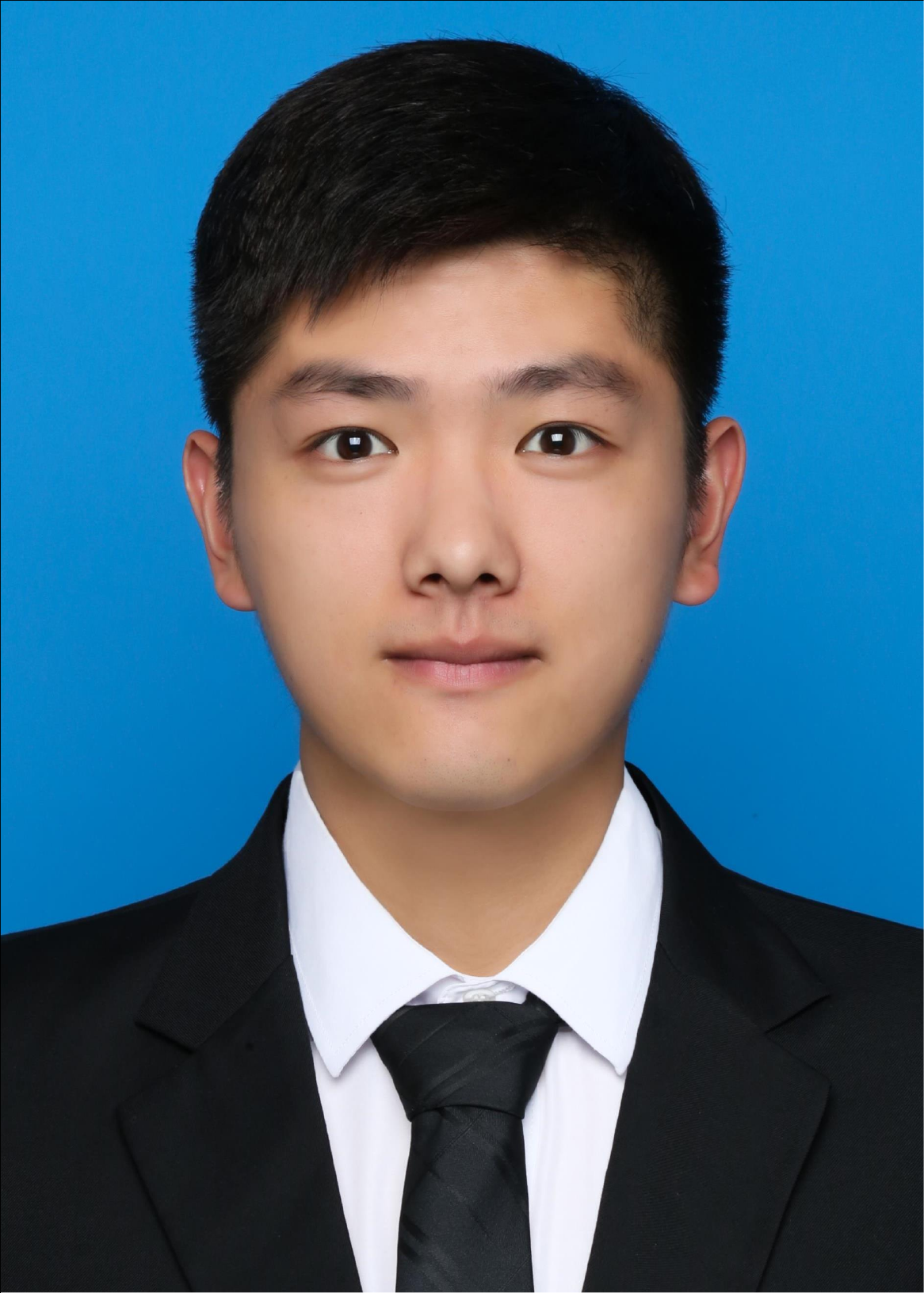}}]{Xiaoxu PENG}
was born in Harbin (China) in 1995.11. received the B.S. and M.S. degrees in College of Mechanical and Electrical Engineering, Northeast Forestry University in 2018 and 2021 respectively, and now a doctoral student at Harbin Institute of Technology. His main research interests are Deep learning and Adversarial examples.
\end{IEEEbiography}
\vspace{-20pt}
\begin{IEEEbiography}
[{\includegraphics[width=1in,height=1.25in,clip,keepaspectratio]{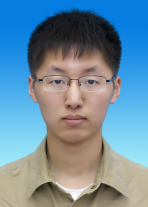}}]{Dong Zhou}
received the B.S degree in automation from Harbin Engineering University, Harbin, China, in 2018. He is currently working toward the Ph.D. degree in the Department of Control Science and Engineering, Harbin Institute of Technology, Harbin, China. His research interests include visual Object Tracking, Adversarial Attack, and Deep Reinforcement Learning. 
\end{IEEEbiography}
\vspace{-20pt}
\begin{IEEEbiography}
[{\includegraphics[width=0.9in,height=1.25in,clip,keepaspectratio]{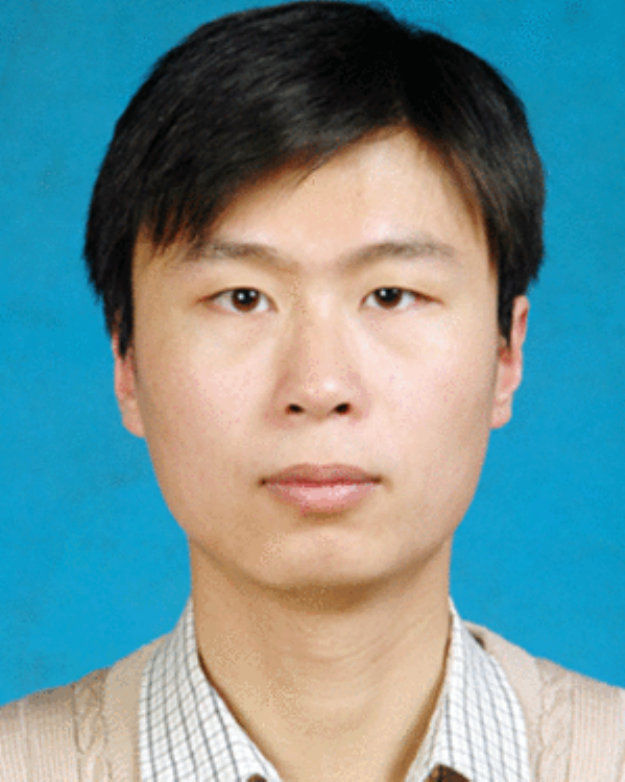}}]{Guanghui SUN}
received the B.S. degree in Automation from Harbin Institute of Technology, Harbin, China, in 2005, and the M.S. and Ph.D. degrees in Control Science and Engineering from Harbin Institute of Technology, Harbin, China, in 2007 and 2010, respectively. He is currently a PROFESSOR with Department of Control Science and Engineering in Harbin Institute of Technology, Harbin, China. His research interests include Machine Learning, Computer Vision, and Adversarial Examples.
\end{IEEEbiography}
\vspace{-20pt}
\begin{IEEEbiography}[{\includegraphics[width=1in,height=1.25in,clip,keepaspectratio]{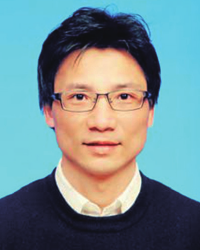}}]{Ligang Wu}(Fellow, IEEE) received the B.S. degree in automation from the Harbin University of Science and Technology, China, in 2001, and the M.E. degree in navigation guidance and control and the Ph.D. degree in control theory and control engineering from the Harbin Institute of Technology, China, in 2003 and 2006, respectively. He was a Research Associate/Senior Research Associate with The University of Hong Kong, the City University of Hong Kong, and Imperial College London. He is currently a Professor with the Harbin Institute of Technology. He has published seven research monographs and more than 200 research articles in internationally referred journals. His current research interests include analysis and design for cyber-physical systems, robotic and autonomous systems, intelligent systems, and power electronic systems. His awards and recognitions include the National Science Fund for Distinguished Young Scholar, and the Highly Cited Researcher since 2015. He also serves as an Associate Editor for a number of journals, including IEEE Transactions on Automatic Control, IEEE Transactions on Industrial Electronics, IEEE/ASME Transactions on Mechatronics, and IET Control Theory and Applications.
\end{IEEEbiography}
\vfill


\end{document}